\documentclass[seceq,preprint]{ptptex}

\usepackage{amsmath,amssymb,latexsym}

\newcommand{\R}{\mathbb R}
\newcommand{\Z}{\mathbb Z}
\newcommand{\RP}{\mathbb{RP}}

\newcommand{\ket}[1]{\left|#1\right>}
\newcommand{\AD}[1]{$\ol{\mbox{D~\,}}\!\!\!$#1}
\newcommand{\Op}[1]{$\mbox{O}#1^+$}
\newcommand{\Om}[1]{$\mbox{O}#1^-$}
\newcommand{\Omt}[1]{$\wt{\mbox{O}#1}^-$}
\newcommand{\Opt}[1]{$\wt{\mbox{O}#1}^+$}
\newcommand{\Omh}[1]{$\wh{\mbox{O}#1}^-$}

\newcommand{\ol}{\overline}

\newcommand{\ra}{\rightarrow}

\newcommand{\cO}{\mathcal O}
\newcommand{\cN}{\mathcal N}

\newcommand{\wt}{\widetilde}
\newcommand{\wh}{\widehat}
\newcommand{\del}{\partial}
\newcommand{\nn}{\nonumber}
\newcommand{\half}{\frac{1}{2}}

\newcommand{\VEV}[1]{\left\langle #1\right\rangle}

\def\Tr{\mathop{\rm Tr}\nolimits}
\def\tr{\mathop{\rm tr}\nolimits}

\def\drawbox#1#2{\hrule height#2pt
        \hbox{\vrule width#2pt height#1pt \kern#1pt
              \vrule width#2pt}
              \hrule height#2pt}

\def\Fund#1#2{\vcenter{\vbox{\drawbox{#1}{#2}}}}
\def\Asymm#1#2{\vcenter{\vbox{\drawbox{#1}{#2}
              \kern-#2pt       
              \drawbox{#1}{#2}}}}
\def\TAsymm#1#2{\vcenter{\vbox{\drawbox{#1}{#2}
              \kern-#2pt       
              \drawbox{#1}{#2}
              \kern-#2pt       
              \drawbox{#1}{#2}}}}

\def\Upbox#1#2{\vcenter{\vbox{\drawbox{#1}{#2}
              \kern+#2pt       
              \drawbox{#1}{0}}}}
\def\TUpbox#1#2{\vcenter{\vbox{\drawbox{#1}{#2}
              \kern+#2pt       
              \drawbox{#1}{0}
              \kern+#2pt       
              \drawbox{#1}{0}}}}

\def\fnd{\Fund{5.5}{0.4}}
\def\asym{\Asymm{5.5}{0.4}}
\def\sym{\fnd\kern-0.4pt\fnd}

\def\mat#1{\matt[#1]}
\def\matt[#1,#2,#3,#4]{\left(%
\begin{array}{cc} #1 & #2 \\ #3 & #4 \end{array} \right)}

\def\vectwoo[#1,#2]{\left(%
\begin{array}{cc} #1 \\ #2 \end{array} \right)}

\notypesetlogo                       
\preprintnumber[3cm]{
IPMU 12-0138}

\title{
Confinement and Dynamical Symmetry Breaking\\
in Non-SUSY Gauge Theory\\ from S-Duality in String Theory
}

\author{
Shigeki \textsc{Sugimoto}%
}

\inst{
Kavli Institute for the Physics and Mathematics of the Universe
(WPI),\\
Todai Institutes for Advanced Study, The University of Tokyo,\\
Kashiwa, 277-8583, Japan
}

\abst{
We discuss an attempt to understand confinement and dynamical symmetry
breaking in a 4-dimensional non-supersymmetric gauge theory using
 S-duality in type IIB string theory. The electric theory is a $USp(2n)$
 gauge theory and its magnetic dual is argued to be an $SO(2n)$ or
$SO(2n-1)$ gauge theory. These theories are obtained as the low-energy
effective theory of O3-\AD3 systems in type IIB string theory, which
are related by S-duality. Confinement and dynamical symmetry
breaking in the $USp(2n)$ gauge theory are caused by the
condensation of scalar fields in the magnetic dual description, which is
consistent with the scenario of the dual Meissner mechanism for the confinement.
}

\begin{document}

\maketitle

\section{Introduction}

Confinement is one of the most profound phenomena in strongly coupled
gauge theory. In many strongly coupled gauge theories including
Yang-Mills theory and quantum chromodynamics (QCD), it is believed that
the elementary particles that belong to non-trivial representations of
the gauge group can never be observed as isolated particles.
An intuitive explanation for the confinement is that
the gauge flux associated with each gauge non-singlet particle
will be squeezed into a thin tube with finite tension and
it costs a huge amount of energy to separate a particle from others,
since the energy is proportional to the length of the flux tube.

Showing the formation of flux tubes for a given gauge theory
is a highly non-trivial problem. Because it is a phenomenon in
strongly coupled systems, perturbative calculations are not reliable.
One scenario for the confinement
proposed in the 1970's \cite{Nambu:1974zg,tHooft,Mandelstam:1974pi} 
is based on the analogy of the Meissner effect in (type II)
superconductors. In superconductors, the $U(1)$ electromagnetic gauge
symmetry is totally Higgsed by the condensation of Cooper pairs
and, as a consequence, the magnetic flux is squeezed into
a flux tube. The idea is to consider the electric-magnetic duality
of the Meissner effect to show the existence of thin flux tubes
associated with electrically charged particles.
Suppose that there is a dual magnetic description
of QCD and the magnetic gauge symmetry is somehow Higgsed by
monopole condensation,
then the electric-magnetic dual of the Meissner mechanism
suggests the formation of a squeezed color flux tube,
which implies the confinement in QCD.
Although this dual Meissner effect is very appealing,
there are a number of questions to be answered.
What is the magnetic dual of QCD? What is the monopole in QCD?
Do the monopoles really condense?
There have been many investigations attempting to clarify these
issues.\footnote{See, for example, Refs.~\citen{Ripka:2003vv}
and \citen{Greensite:2011zz} for reviews.}

In the mid 1990's, it was shown that this scenario of confinement is
realized in some examples of supersymmetric gauge theories.
In Ref.~\citen{Seiberg:1994rs}, the low-energy effective theory
of the $\cN=2$ $SU(2)$ super Yang-Mills (SYM) theory
was exactly obtained and it was shown that the magnetic monopoles
become massless at a singularity of the vacuum moduli space.
Furthermore, when a mass term for the chiral multiplet that breaks
$\cN=2$ supersymmetry to $\cN=1$ is added, the magnetic $U(1)$ gauge
symmetry is shown to be Higgsed via the condensation of the magnetic
monopoles. This is consistent with the conjecture that the $\cN=1$
$SU(2)$ SYM is a confining theory.
This argument was soon generalized to a much wider class of
supersymmetric gauge theories.
In Ref.~\citen{Seiberg:1994pq}, a non-Abelian generalization of the
electric-magnetic duality in the $\cN=1$ supersymmetric QCD (SQCD)
was found.
The ``electric theory'' in Seiberg's duality is the $\cN=1$ $SU(N_c)$
SQCD with $N_f$ flavors and the ``magnetic theory'', which is
conjectured to be the magnetic dual of the electric theory, is the
$\cN=1$ $SU(N_f-N_c)$ SQCD with $N_f$ flavors with an additional gauge
singlet field. It was argued that when one of these theories
is completely Higgsed by the condensation of the scalar components of
the quark superfields, the other theory is confined.
This observation suggests that an analogue of the dual Meissner
mechanism for the confinement also works for the cases where
both electric and magnetic descriptions are non-Abelian gauge theories.

In this paper, we investigate an example of duality in
non-supersymmetric gauge theories.
As we will describe in \S \ref{sec2},
the electric theory is a $USp(2n)$ gauge theory
and the magnetic theory is an $SO(2n)$ gauge theory with a tachyonic
scalar field. The magnetic $SO(2n)$ gauge theory will flow to an
$SO(2n-1)$ gauge theory after tachyon condensation.
In general, collecting convincing non-trivial evidence for the duality
in non-supersymmetric gauge theory is much more difficult than in
supersymmetric gauge theory.\footnote{See
Refs.~\citen{Schmaltz:1998bg,Armoni:2008gg,Mojaza:2011rw} for some examples.}
Our model is based on the S-duality of O3-\AD3
systems in type IIB string theory, which was first analyzed by Uranga
in Ref.~\citen{Uranga:1999ib}. 
Although the proof is not yet available, the S-duality in type IIB
string theory is considered to be an exact duality, which should
hold even in non-supersymmetric situations.
Since our gauge theories are realized as the low-energy effective theories
of the O3-\AD3 systems, we can apply the S-duality in string theory
and investigate its consequences in these non-supersymmetric
gauge theories.
One of the main goals of this paper is to propose a scenario to
understand the confinement and dynamical symmetry breaking expected in the
electric theory using the magnetic description as well as
some knowledge of string theory.

Even though we have some techniques in string theory
to analyze the properties of the gauge theories, it is still not easy
to obtain exact results in the non-supersymmetric gauge theories,
and hence our arguments in this paper are mostly qualitative
and sometimes speculative.
In particular, at the energy scale of the dynamical symmetry breaking,
neither electric nor magnetic theories are weakly coupled and
we have to rely on a toy model that seems to capture the qualitative
features of the systems to gain a consistent picture.
Therefore, we are {\it not} going to prove the confinement
and/or dynamical symmetry breaking, but at best try to understand
what is going on under the duality.
Nevertheless, we hope that this analysis will shed some light on
the confinement problem in non-supersymmetric gauge theories.

The paper is organized as follows.
In \S \ref{sec2}, we start with describing the
electric and magnetic theories.
In \S \ref{sec3}, we review some of the known facts
about the O3-planes and their S-duality properties,
and explain how we obtain our electric and magnetic theories
realized on the O3-\AD3 systems.
The confinement and dynamical symmetry breaking
in our systems are studied in \S \ref{sec4}.
In \S \ref{sec5}, we summarize what we have revealed and discuss some
problems to be solved.

We have tried to make the paper readable for people who are not
familiar with string theory. Although we need to use
knowledge from string theory to explain the logic behind
the duality, our proposal and its consequences can be followed without
using string theory. Those who want to avoid
some details involving string theory can skip
\S \ref{sec3}, and go directly to \S \ref{sec4} after reading
\S \ref{sec2}. Section \ref{fluxtube} can also be skipped.

\section{Electric and magnetic theory}
\label{sec2}

Here, we describe the low-energy effective theory of the electric and
magnetic theories, which are conjectured to be S-dual to each other.
As we will see in detail in \S \ref{sec3}, both electric and
magnetic theories are realized in type IIB string theory, and the duality
between them follows from the well-known S-duality in string theory.
In this section, in order to present our proposal clearly, we write
down the low-energy field contents of the electric and magnetic theories,
and discuss some of the supporting evidence of the duality,
which can be easily seen without going into detail.

\subsection{Electric theory}
\label{eleth}
The low-energy field content of the electric theory is summarized in
Table \ref{ele}.
\begin{table}[hb]
\begin{eqnarray}
 \begin{array}{c|cc}
 &USp(2n) &SO(6) \\
\hline
A_\mu &  \sym &1\\
Q^i & \asym& ~\,4_+\\
\Phi^I & \sym&6\\
 \end{array}
\nn
\end{eqnarray}
\caption{Electric theory.}
\label{ele}
\end{table}
Here, the unitary symplectic group $USp(2n)$ is the gauge
group and $SO(6)$ is the global symmetry.\footnote{
See Appendix \ref{USp} for our notation of the $USp(2n)$ gauge group.}
$A_\mu$, $Q^i$ and $\Phi^I$ in Table \ref{ele} are the gauge field,
left-handed Weyl fermions and Hermitian scalar fields, respectively.
The rank-2 symmetric tensor representation ($\sym$) of the gauge group
$USp(2n)$ is equivalent to the adjoint representation.
In our notation, the rank-2 antisymmetric tensor representation
($\asym$) is a reducible representation and the irreducible
components consist of the singlet representation and
its orthogonal complement.
(See Appendix \ref{USp} for more detail.)

The notations ``$4_+$'' and ``$6$'' in Table \ref{ele} stand for the
positive chirality spinor representation and the 6-dimensional vector
representation of $SO(6)$, respectively.
The index $i$ ($i=1,2,3,4$) for $Q^i$ is the $4_+$ spinor index for
$SO(6)$, which can be thought of as the fundamental
representation of $SU(4)\simeq SO(6)$. The index
$I$ ($I=1,2,\cdots,6$) for $\Phi^I$ is the vector index of $SO(6)$.
Note that if $Q^i$ were in the adjoint representation ($\sym$) of the
gauge group $USp(2n)$, the electric theory would be the $\cN=4$
$USp(2n)$ SYM.

This theory is obtained by imposing the conditions
\begin{eqnarray}
(JA_\mu)^T = JA_\mu\ ,~~
(JQ^i)^T = -JQ^i\ ,~~
(J\Phi^I)^T = J\Phi^I
\label{JAJQJP}
\end{eqnarray}
for the gauge field, fermion and scalar fields, respectively,
in the $\cN=4$ $SU(2n)$ SYM.

Then, the tree level Lagrangian density is schematically written as
\begin{eqnarray}
{\cal L}^{\rm tree}_{\rm electric}&\sim&\Tr F_{\mu\nu}^2
+\Tr \left(\bar Q_i\sigma^\mu(\del_\mu Q^i+[A_\mu, Q^i])\right)
+\Tr\left((\del_\mu\Phi^I+[A_\mu,\Phi^I])^2\right)
\nn\\
&&
+\Tr\left(Q^i\Sigma^I_{ij}[\Phi^I, Q^j]\right)
+\Tr\left([\Phi^I,\Phi^J]^2\right)+{\rm h.c.}
\ ,
\label{Lele}
\end{eqnarray}
where $\bar Q_i$ is the Hermitian conjugate of $Q^i$
and $\Sigma_{ij}^I$ is the invariant tensor in the
$(4_-\otimes 4_-)_{\rm asym}\otimes 6$ representation of $SO(6)$.
(See Appendix \ref{Sigma} for the explicit form.)
Here all the couplings are omitted.

The one-loop beta function for the gauge coupling indicates that
this theory is asymptotically free, and hence the electric theory
is better described at high energies.
All the fields in Table \ref{ele} are massless at the tree level.
However, since the supersymmetry is completely broken, there is no
reason for the scalar fields to remain massless after taking into
account the quantum effect. In fact, the explicit one-loop calculation
implies that all the scalar fields will acquire masses on the order
of the cutoff scale, which is the string scale, and will be decoupled
at low energy. (See \S \ref{oneloopmass} for more details.)
According to the recent analysis in
Refs.~\citen{Sannino:2009aw,Armoni:2009jn,Golkar:2009aq},
the $USp(2n)$ theory with four Weyl fermions in the antisymmetric
representation is conjectured to be outside the conformal window
and is in the confined phase. Furthermore, for the $n>1$ cases,
the global $SO(6)$ symmetry is expected to be
dynamically broken to the $SO(4)$ subgroup via
condensation of the fermion bilinear operator:
\begin{eqnarray}
\epsilon^{\alpha\beta} \langle \Tr(Q_\alpha^iQ_\beta^j)
\rangle \propto \delta^{ij}\ ,
\label{TrQQ}
\end{eqnarray}
where $\alpha,\beta$ are the Lorentz spinor indices.
Note that the indices $i,j$ in (\ref{TrQQ}) are those of
the fundamental representation of $SU(4)\simeq SO(6)$
and the unbroken subgroup $SO(4)$ is the real part
of $SU(4)$.

Associated with this symmetry breaking,
we expect to have Nambu-Goldstone modes with values in the coset space
$SU(4)/SO(4)$. In common with pions in QCD, the Nambu-Goldstone particle
is provided as a bound state of two fermions.
The other particles in the spectrum are expected to be massive,
except for the gauge singlet part of the fermions $Q^i$,
because there is no symmetry to protect the mass terms for them.

The $n=1$ case is also interesting. The gauge group for $n=1$ is
$USp(2)\simeq SU(2)$ and since the antisymmetric representation
is a gauge singlet,
the fermions will be decoupled from the gauge field.
Therefore, the low energy dynamics of the electric theory
is expected to be equivalent to the $SU(2)$ pure Yang-Mills theory.
Again, this theory is conjectured to be a confining theory,
although the $SO(6)$ symmetry will not be broken in this case.

\subsection{Magnetic theory}
\label{magth}

The low-energy field content of the magnetic theory is given in
Table \ref{mag1}.
\begin{table}[hb]
\begin{eqnarray}
 \begin{array}{c|cc}
&SO(2n) &SO(6) \\
\hline
a_\mu & \asym &1\\
q^i & \sym& ~\,4_+\\
\phi^I & \asym&6\\
t&\fnd&1\\
\psi_i&\fnd& ~\,4_-
 \end{array}
\nn
\end{eqnarray}
\caption{Magnetic theory (I).}
\label{mag1}
\end{table}
Here, we have listed the fields that are massless or tachyonic
at the tree level. The gauge group is $SO(2n)$ and $a_\mu$ is
the associated gauge field. $q^i$ and $\psi_i$ ($i=1,\cdots,4$)
are left-handed Weyl fermions, $\phi^I$ ($I=1,\cdots,6$)
are Hermitian scalar fields and $t$ is a tachyon field,
which is a real scalar field with negative mass squared.
$\asym$ , $\sym$ and $\fnd$ in Table \ref{mag1} are
the rank-2 antisymmetric tensor representation,
rank-2 symmetric tensor representation and vector representation
of $SO(2n)$, respectively. Note that
the rank-2 symmetric tensor representation
($\sym$) is a reducible representation, and 
the irreducible components are
the trace part and the traceless part.

The tree-level Lagrangian contains the following terms:
\begin{eqnarray}
{\cal L}^{\rm tree}_{\rm magnetic\, (I)}
&\sim&\tr f_{\mu\nu}^2
+\tr \left(\bar q_i\sigma^\mu(\del_\mu q^i+[a_\mu, q^i])\right)
+\tr\left((\del_\mu\phi^I+[a_\mu,\phi^I])^2
\right)
\nn\\
&&+
\left((\del_\mu+ a_\mu) t \right)^2+V(t)
+\bar\psi^i\sigma^\mu(\del_\mu \psi_i+a_\mu \psi_i)
\nn\\
&&+
\tr\left(q^i\Sigma^I_{ij}[\phi^I, q^j]\right)
+\tr\left([\phi^I,\phi^J]^2\right)
\nn\\
&&
+ t^T\,\phi^I\phi^I t
+\bar\psi^{iT}\Sigma^{I}_{ij}\phi^I\bar\psi^j
+t^T\,q^i\psi_i+{\rm h.c.}
\label{Lmag1}
\end{eqnarray}
Again, this is a schematic expression and
 all the couplings as well as possible higher dimensional
terms are omitted.
A sketch of the derivation of this Lagrangian will be given
in \S \ref{O3D3bar}.
Here, $V(t)$ is the potential for the tachyon field $t$.
Since we are mostly interested in the qualitative properties,
we do not need the explicit form of the potential.
The only property we need here is that the potential $V(t)$ is
unstable around the origin $t=0$ and hence the tachyon field
will develop a non-trivial vacuum expectation value.
We refer to this theory as magnetic theory (I).

After tachyon condensation, the gauge symmetry will be broken to
$SO(2n-1)$. The terms $t^T\phi^I\phi^I t$ and $t^T q^i\psi_i$ in the
last line of (\ref{Lmag1}) imply that $\psi^i$ as well as some
components of $\phi^I$ and $q^i$ will become massive. The massless
components obtained after tachyon condensation are
listed in Table \ref{mag2}.
\begin{table}[hb]
\begin{eqnarray}
 \begin{array}{c|cc}
 &SO(2n-1) &SO(6) \\
\hline
a_\mu  & \asym &1\\
q^i  & \sym& ~\,4_+\\
\phi^I  & \asym&6\\
 \end{array}
\nn
\end{eqnarray}
\caption{Magnetic theory (II).}
\label{mag2}
\end{table}
The effective Lagrangian for the massless fields after tachyon
condensation will be
\begin{eqnarray}
{\cal L}^{\rm tree}_{\rm magnetic\, (II)}&\sim&\tr f_{\mu\nu}^2
+\tr \left(\bar q_i\sigma^\mu(\del_\mu q^i+[a_\mu, q^i])\right)
+\tr\left((\del_\mu\phi^I+[a_\mu,\phi^I])^2
\right)
\nn\\
&&
+\bar\psi^i\sigma^\mu(\del_\mu \psi_i+a_\mu \psi_i)
+\tr\left(q^i\Sigma^I_{ij}[\phi^I, q^j]\right)
+\tr\left([\phi^I,\phi^J]^2\right)+{\rm h.c.}
\label{Lmag2}
\end{eqnarray}
This Lagrangian can be obtained by imposing
\begin{eqnarray}
A_\mu^T=-A_\mu\ ,~~~
q^{i\,T}=q^i\ ,~~~
\phi^{I\,T}=-\phi^I\ ,
\label{AQPproj}
\end{eqnarray}
in the $\cN=4$ SYM with the gauge group $SU(2n-1)$.
We refer to this theory as magnetic theory (II).

\subsection{First view on the duality}
\label{first}

As we will see in more detail in the following sections, we claim that
the electric theory in \S \ref{eleth} and
the magnetic theory described in \S \ref{magth}
are dual to each other.
Or, more precisely, they are low-energy effective theories
of the two brane configurations related by S-duality
in type IIB string theory.
In this subsection, we make a few comments on the duality, which
can be seen without knowing the detailed structure.

As a first check, one can easily see that the 't Hooft anomaly matching
condition with respect to the global $SO(6)$ symmetry is
satisfied, as pointed out in Ref.~\citen{Uranga:1999ib}
for the $n=1$ case.\footnote{Here, we consider the $SO(6)^3$ anomaly
by carrying out our analysis
as if the $SO(6)$ symmetry were not broken. However,
the $SO(6)$ symmetry is actually expected to be spontaneously broken,
as we have discussed in \S \ref{eleth}. Then, the same amount of
anomaly is induced by the associated Nambu-Goldstone bosons with the
Wess-Zumino-Witten term.}
In fact, the $SO(6)^3$ anomaly in the electric theory
is proportional to $n(2n-1)$ and the corresponding
quantity in magnetic theory (I) is
$n(2n+1)-2n=n(2n-1)$,
where $n(2n+1)$ and $-2n$ are the contributions
from $q^i$ and $\psi_i$ in Table \ref{mag1}, respectively.
The result for magnetic theory (II) is also $n(2n-1)$.

One may wonder whether this claim is consistent with
Goddard-Nuyts-Olive (GNO) duality \cite{Goddard:1976qe},
in which the dual group of $USp(2n)$ is argued to be $SO(2n+1)$.
At the tree level, the bosonic part of the electric theory
is the same as the $\cN=4$ $USp(2n)$ SYM
and there are flat directions for the scalar fields $\Phi^I$.
At a generic point along the flat directions, the gauge group
is broken to $U(1)^n$ and the monopole solutions can be constructed.
Although the flat directions will be lifted by quantum corrections,
it is natural to expect that these monopoles are related to the
fields in the magnetic theory.
If this is the case, the monopoles should satisfy the Dirac quantization
conditions and the argument of GNO should be applied.
In fact, what GNO showed is that the magnetic charges
take values in the weight lattice of the dual group,
which is satisfied for the fields in Table \ref{mag1},
if we regard the gauge group $SO(2n)$ in magnetic theory (I)
as a subgroup of the GNO dual group $SO(2n+1)$.
However, the existence of the massless gauge particle associated with the dual
group in the spectrum is not guaranteed in general,
and hence, there is no contradiction with the argument of GNO,
even though the gauge group in our magnetic theory is not $SO(2n+1)$.

Since both the electric and magnetic theories are closely related to the
$\cN=4$ SYM, the well-known duality in the $\cN=4$ SYM appears in our
system in various ways. For example, if we consider a configuration with
\begin{eqnarray}
\Phi^1=v\mat{1_n,0,0,-1_n}
\label{Phi1}
\end{eqnarray}
in the electric theory, the gauge group $USp(2n)$ will be broken to
$U(n)$, and then the system can be seen as the $\cN=4$ $U(n)$ SYM
coupled with massive non-supersymmetric fields.
Although the configuration (\ref{Phi1}) is quantum mechanically
unstable, because the mass term for the scalar fields $\Phi^I$ will be
generated by quantum corrections, let us assume here that considering
the theory around this configuration makes sense in some limit;
for example, the large $n$ limit discussed below or the large $v$ limit
in the string theory setup.
On the other hand, this configuration corresponds to setting
\begin{eqnarray}
 \phi^1=iv\mat{0,1_n,-1_n,0}
\label{magphi1}
\end{eqnarray}
in magnetic theory (I).
Then, the $SO(2n)$ gauge group is broken to $U(n)$ and, if $v$ is large
enough, the tachyon will become massive and the massless degrees of
freedom again give the $\cN=4$ $U(n)$ SYM, which is the magnetic dual of
the $\cN=4$ $U(n)$ SYM appeared in the electric theory.
In the string theory realization explained in \S \ref{sec3},
the configurations (\ref{Phi1}) in electric theory
and (\ref{magphi1}) in magnetic theory (I)
correspond to the configuration with $n$ \AD3-branes placed away from
the \Op3-plane and \Omt3-plane, respectively, and the $\cN=4$ $U(n)$ SYM
is obtained from the open strings attached on the \AD3-branes.

The large $n$ behavior is also related to the $\cN=4$ SYM.
Note that if we replace the fermions in the electric theory and
magnetic theory (II) with those in the adjoint representation of the
gauge group, these theories will become the $\cN=4$ SYM.
Because the difference between the contributions of the
fields in the symmetric and antisymmetric tensor representations
is subleading in the $1/n$ expansion, both the electric and magnetic
theories are equivalent to the $\cN=4$ SYM at the leading order of the
$1/n$ expansion.\footnote{See
Refs.~\citen{Kovtun:2004bz,Armoni:2004ub,Unsal:2006pj,Armoni:2007jt}
for more precise statements and rigorous arguments.}
Therefore, at the leading order of the $1/n$ expansion,
the duality of our electric and magnetic theories
is related to that of the $\cN=4$ SYM.

\section{S-duality of O3-planes and O3-\AD3 systems}
\label{sec3}

In this section, we briefly review some properties of the O3-planes
and O3-\AD3 systems that will be used in the following discussion.
The main references are
Refs.~\citen{Witten:1998xy}, \citen{Uranga:1999ib} and
\citen{Hyakutake:2000mr}.
Readers who want to avoid string theory can skip this section
and go directly to \S \ref{sec4}.

\subsection{O3-planes and S-duality}
\label{O3S}

We consider 10-dimensional flat space-time parametrized by
$x^0,x^1,\cdots,x^9$ in type IIB string theory with an O3-plane localized
at $x^4=x^5=\cdots=x^9=0$.
The O3-plane is defined as the (3+1)-dimensional fixed plane
with respect to the $\Z_2$ orientifold action generated
by $I_6\Omega(-1)^{F_L}$. Here $I_6$ flips the sign of six spatial
coordinates $x^{4\sim 9}$ transverse to the O3-plane,
$\Omega$ is the world-sheet parity transformation and $F_L$ is the
left-moving space-time fermion number.\footnote{See,
for example, Ref.~\citen{Dabholkar:1997zd} for a review of orientifolds.}
There are at least four types of O3-planes, denoted as \Om3, \Op3,
\Omt3, \Opt3,
depending on the choice of the discrete torsions associated with NSNS and
RR 2-form fields.\cite{Witten:1998xy}
The discrete torsions are defined as
\begin{eqnarray}
\tau_{\rm NS}=\exp\left(
i\int_{\RP^2} B_2
\right)\ ,~~~
\tau_{\rm RR}=\exp\left(
i\int_{\RP^2} C_2
\right)\ ,
\end{eqnarray}
where $B_2$ and $C_2$ are the NSNS and RR 2-from fields, respectively,
and ${\RP^2}$ is a two-sphere surrounding the O3-plane
in the $x^{4\sim 9}$-plane ($\R^6$ parametrized by $x^{4\sim 9}$)
divided by the $\Z_2$ orientifold action.
$\tau_{\rm NS}$ and $\tau_{\rm RR}$ can take values in $\{\pm 1\}$.
 \Om3, \Op3, \Omt3, \Opt3 are the O3-planes with $(\tau_{\rm NS},\tau_{\rm
 RR})=(+,+), (-,+), (+,-), (-,-)$, respectively.
Recall that the type IIB string theory is believed to be invariant under
the action of the S-duality group $SL(2,\Z)$, which acts
on $B_2$ and $C_2$ as
\begin{eqnarray}
\left({C_2\atop B_2}\right)\ra
\Lambda\left({C_2\atop B_2}\right)\ ,~~\Lambda\in SL(2,\Z)\ .
\label{SonCB}
\end{eqnarray}
It also acts on the dilaton field $\phi$ and RR 0-form field $C_0$ as
\begin{eqnarray}
 \tau\ra \frac{a\tau+b}{c\tau+d}\ ,~~~\Lambda=\mat{a,b,c,d}\in SL(2,\Z)\ ,
\end{eqnarray}
where $\tau=C_0+i e^{-\phi}$ is a combination of the RR 0-form field $C_0$
and the dilaton $\phi$.
We are particularly interested in the action of
\begin{eqnarray}
S\equiv \mat{0,-1,1,0}\in SL(2,\Z)\ ,
\label{S}
\end{eqnarray}
which gives the strong/weak duality, since
it acts on the string coupling $g_s=e^{\phi}$ as
$g_s\ra 1/g_s$.
In this paper, we use the term ``S-duality'' for the action of $S$
in (\ref{S}), rather than the full $SL(2,\Z)$ action.
{}From (\ref{SonCB}), we can easily read how the $SL(2,\Z)$ acts on the
discrete torsion $(\tau_{\rm NS},\tau_{\rm RR})$.
In particular, the \Omt3-plane and the \Op3-plane are interchanged under
the S-duality, while the \Om3-plane is S-duality invariant.
The \Opt3-plane is related to the \Op3-plane by the T-transformation
given by the action of
\begin{eqnarray}
T\equiv\mat{1,1,0,1}\in SL(2,\Z)\ .
\end{eqnarray}

Let us next consider a system with $n$ D3-branes placed at the O3-plane.
The low-energy effective theory of the open strings attached
on the D3-brane is the $\cN=4$ SYM.
The gauge groups for the systems with \Om3, \Op3, \Omt3, \Opt3-planes
are $SO(2n)$, $USp(2n)$, $SO(2n+1)$, $USp(2n)$, respectively.
The gauge coupling $e$ and the theta parameter $\theta$ are
related to the dilaton and RR 0-form field $C_0$ as\footnote{
The normalization of the gauge coupling and theta parameter
is different from that used, for example, in Ref.~\citen{Kapustin:2006pk}.
Here, the period of $\theta$ is $2\pi$ for $SO(N)$ and $4\pi$
for $USp(2n)$ theory. This normalization is natural in string theory,
since a half D-instanton (one D-instanton before the orientifold
projection) is allowed for the \Op3-plane, while only an integer number
of D-instantons is allowed for the \Om3-plane.
}
\begin{eqnarray}
 \tau=C_0+i e^{-\phi}=\frac{\theta}{2\pi}+\frac{4\pi i}{e^2}\ .
\end{eqnarray}
The $SL(2,\Z)$ duality in the string theory considered above is consistent
with the $SL(2,\Z)$ duality in the $\cN=4$ SYM
with these gauge groups \cite{Witten:1998xy}.
In particular, the $\cN=4$ SYM with the $USp(2n)$ gauge group is S-dual
to that with the $SO(2n+1)$ gauge group,
which is consistent with the fact that
$\tau_{\rm NS}$ and $\tau_{\rm RR}$ are interchanged under the S-duality.

The perturbative spectrum of the system with the \Omt3-plane and
$n$ D3-branes is formally the same as that for the \Om3-plane with
$(n+1/2)$ D3-branes, that is, $(2n+1)$ D3-branes on the covering space
before the orientifold projection.\footnote{
In some literature, such as, Ref.~\citen{Uranga:1999ib}, the number of
D-branes is counted on the covering space before the orientifold
projection, which is double the number given by our convention.}
Therefore, the \Omt3-plane can be regarded as the \Om3-plane with a half
D3-brane stuck on it, at least in perturbative calculations.
However, this picture is misleading in non-perturbative analysis.
If we naively consider \Omt3-plane as the \Om3-plane with a half D3-brane
and apply the S-duality to the \Om3-plane and D3-brane components,
the \Omt3-plane may appear S-duality invariant, since both the
\Om3-plane and D3-brane
are S-duality invariant. However, as we have seen above, the \Omt3-plane
is mapped to the \Op3-plane and vice versa under the S-duality.
Another important property that cannot be understood in this
naive picture is that D-strings (D1-branes) cannot have their end points
on the \Omt3-plane, although fundamental strings can be attached on the
\Omt3-plane. {}From the perturbative open string spectrum, it can be
shown that the fundamental strings can be attached on the \Omt3-plane
but not on the O$3^\pm$-planes. Since a fundamental string stretched
along the radial direction of the $x^{4\sim 9}$-plane is a BPS object,
we expect that this property will hold even for the strongly coupled
regime.
Applying the S-duality, we see that D-strings can be attached on the
\Op3-plane, but not on the \Om3 or \Omt3-planes.

These properties can be better understood from the construction
of the \Omt3 and \Op3-planes given in Ref.~\citen{Hyakutake:2000mr}.
It was argued that the \Omt3-plane can be continuously deformed
to a system with an \Om3-plane and a spherical D5-brane surrounding it.
Here, the world-volume of the spherical D5-brane has a topology
of $\R^{1,3}\times S^2/\Z_2$, where $\R^{1,3}$ corresponds to the
directions parallel to the \Om3-plane,
$S^2$ is a two-sphere surrounding the \Om3-plane in the
$x^{4\sim 9}$-plane
and $\Z_2$ is the orientifold action acting on it.
Because the sphere should be invariant under the
$\Z_2$ action, the spherical D5-brane cannot escape from the sphere
to infinity.
Without any other fluxes,
the radius of the sphere will shrink to zero size and
the $SO(6)$ rotational symmetry will be restored, but its
vestiges will be retained in the RR charge and the discrete torsion.
In fact, it was shown that the spherical D5-brane induces the
discrete torsion $(\tau_{\rm NS},\tau_{\rm RR})=(+,-)$, and
carries a half unit of magnetic flux,
\begin{eqnarray}
 \frac{1}{2\pi}\int_{\RP^2} F=\half\ ,
\label{magflux}
\end{eqnarray}
that provides the RR charge of a half D3-brane stuck
on it.\cite{Freed:1999vc,Aharony:2000cw,Hyakutake:2000mr}
Similarly, the \Op3-plane is obtained as an \Om3-plane with a
spherical NS5-brane surrounding it.
Since a D5-brane is mapped to an NS5-brane under the S-duality,
this picture is consistent with the S-duality.

Furthermore,
note that fundamental strings can end on the D5-brane but not on the
NS5-brane, and similarly  D-strings can end on the NS5-brane
but not on the D5-brane.\cite{Strominger:1995ac}${}^{,}$\footnote{
It may be possible to consider a D-string absorbed in the D5-brane
world-volume as an instanton-like gauge configuration
\cite{Witten:1995gx,Douglas:1995bn,Nekrasov:1998ss} or a bound state of
the D1-D3-D5 branes considered in Ref.~\citen{Polchinski:2000uf}.
We will not regard such configurations as those with a D-string ending
on D5-branes, since the D-string still behaves as a string embedded in
the D5-brane world-volume as the energy is proportional to its length.}
These facts suggest that the fundamental strings can be attached on the
\Omt3-plane, but not on the O$3^\pm$-planes and
D-strings can be attached on the \Op3-plane but not on the \Om3 or
\Omt3-planes.

When the size of the sphere of the D5-brane is smaller than the string
length scale, the argument based on the D5-brane world-volume gauge
theory cannot be justified, since the D5-brane world-volume would be
highly curved. However, the properties of the O3-planes discussed above
remain valid even for the zero-size limit, which suggests a continuity
in the size of the sphere of the D5-brane surrounding the O3-plane. This
interpretation of the \Omt3-plane will be used as a useful guide for
understanding the properties of the \Omt3-plane and related objects
in the following sections. 

\subsection{O3-\AD3 systems}
\label{O3D3bar}

Here, we consider O3-\AD3 systems obtained by
rotating the D3-branes considered in the previous
subsection by 180 degrees. We refer to the rotated D3-branes as
\AD3-branes (anti-D3-branes). Since the orientation of the \AD3-brane
is opposite to those considered in the previous subsection,
the supersymmetry preserved by the \AD3-branes is opposite
to that preserved in the existence of the O3-plane.
Therefore, supersymmetry is completely broken in the O3-\AD3 systems.
As analyzed in Refs.~\citen{Sugimoto:1999tx}
and \citen{Uranga:1999ib},\footnote{
See Ref.~\citen{Angelantonj:2002ct} for a review.}
the spectrum of the open strings attached on the O$3^\pm$-\AD3 system
is obtained by simply replacing the fermions in
the symmetric and antisymmetric representations of the gauge group
for the O$3^\pm$-D3 system with those in
the antisymmetric and symmetric representations, respectively.

The electric theory described in \S \ref{eleth} is obtained by
putting $n$ \AD3-branes on top of the \Op3-plane. The massless fields on
the \AD3-branes are as listed in Table \ref{ele}.
The global $SO(6)$ symmetry corresponds to the rotational symmetry in
the $x^{4\sim 9}$-plane. All the other fields have string scale masses.
Without the $\Z_2$ orientifold projection associated with the
\Op3-plane, the system is supersymmetric, and the low-energy
effective theory on the \AD3-branes is the $\cN=4$ $SU(2n)$ SYM.
Therefore, the tree level Lagrangian for the massless fields
(\ref{Lele}) is obtained by imposing the orientifold projection
(\ref{JAJQJP}) in the $\cN=4$ $SU(2n)$ SYM.

The magnetic theory is obtained by applying the S-duality to the
\Op3-\AD3 system. As reviewed in the previous subsection,
the \Op3-plane is mapped to the \Omt3-plane under the S-duality,
while \AD3-branes are S-duality invariant.
Therefore, the S-dual of the electric theory is given by the
system with $n$ \AD3-branes on top of the \Omt3-plane
\cite{Uranga:1999ib}, which we refer to as magnetic theory (I).
The perturbative open string spectrum can be analyzed by
regarding the \Omt3-plane as an \Om3-plane with a half D3-brane.
The massless spectrum of the particles created by open strings with both
end points attached on the \AD3-branes is obtained by replacing
the fermions in the $\cN=4$ $SO(2n)$ SYM with those with a symmetric
representation of the gauge group. The gauge field $a_\mu$, Weyl
fermions $q^i$ and real scalar fields $\phi^I$ in Table \ref{mag1}
are obtained in this way.
In addition, there are strings stretched between one of the \AD3-branes
and the half D3-brane stuck at the \Om3-plane.
The spectrum of this type of open strings is given by the opposite
GSO projection in the NS-R formulation,\cite{Sen:1998ii}$^,$\footnote{
See Ref.~\citen{Sen:1999mg} for a review.} and we obtain
the tachyon field $t$ and the massless fermions $\psi_i$
in Table \ref{mag1}.

As the existence of the tachyon field shows, magnetic theory (I)
is unstable.
The tachyon field will eventually roll down the potential and condense.
Magnetic theory (II) is defined as the low-energy effective theory
obtained via tachyon condensation in magnetic theory (I).
It is now well-established that tachyon condensation
in a system with a D3-\AD3 pair corresponds to
the annihilation of the D3-\AD3
pair.\cite{Sen:1998ii,Sen:1998sm}${}^{,}$\footnote{See also
Ref.~\citen{Sen:1999mg} for a review.}
As proposed in Ref.~\citen{Uranga:1999ib}, it is then natural to expect that
the \Omt3-plane with a \AD3-brane will become an \Om3-plane with a half
\AD3-brane stuck on it after tachyon condensation.
We refer to this object as an \Omh3-plane. In terms of the interpretation
of the \Omt3-plane as an \Om3-plane with a spherical D5-brane explained in
\S \ref{O3S}, the \AD3-brane will be absorbed in the spherical D5-brane
and the magnetic flux induced on it will be decreased from
(\ref{magflux}) by one unit. Then, the \Omh3-plane can be regarded
as an \Om3-plane with a spherical D5-brane that carries minus half a
unit of magnetic flux,
\begin{eqnarray}
 \frac{1}{2\pi}\int_{\RP^2} F=-\half\ ,
\label{magflux2}
\end{eqnarray}
which gives the RR-charge of a half \AD3-brane.
In general, an \Omt3-plane with $n$ \AD3-branes is expected to become
an \Omh3-plane with $(n-1)$ \AD3-branes after tachyon condensation.
The perturbative spectrum is given by regarding
this system as one with an \Om3-plane and $(n-1/2)$ \AD3-branes.
Then, the massless spectrum of this system is as listed
in Table \ref{mag2}. 

The tree-level Lagrangians for magnetic theory (I)
 (\ref{Lmag1}) and magnetic theory (II) (\ref{Lmag2})
are obtained as follows.
The terms without $t$ and $\psi_i$ are obtained
by imposing the orientifold projection (\ref{AQPproj}) in
the $\cN=4$ SYM with the gauge groups $SU(2n)$ and $SU(2n-1)$ for 
magnetic theories (I) and (II), respectively,
for the same reason as the electric theory explained above.
For magnetic theory (I), we add the terms including
$t$ and $\psi_i$ that are consistent with the symmetry.
In fact, the terms $t^T\phi^I\phi^I t$ and $t^Tq^i\psi_i$
in (\ref{Lmag1}) are needed to ensure that the tachyon condensation
corresponds to the annihilation of a D3-\AD3 pair and
implies the massless field content listed in Table \ref{mag2}.
When \AD3-branes are far from the \Omt3-plane,
the particles created by the open strings stretched
between the \Omt3-plane and \AD3-brane become massive.
Therefore, $t$ and $\psi_i$ become massive when
$\phi^I$ is given by (\ref{magphi1}) with large enough $v$.
The terms $t^T\phi^I\phi^I t$ and
$\bar\psi^{iT}\Sigma^{I}_{ij}\phi^I\bar\psi^j$ in (\ref{Lmag1})
are needed to realize this.
We will not try to fix the coupling constants,
although in principle they can be calculated in string theory.

\subsection{On the decoupling limit}
\label{decoupling}

In string theory, there are infinitely many massive fields
in the spectrum and infinitely many higher dimensional terms
in the Lagrangians (\ref{Lele}), (\ref{Lmag1}) and (\ref{Lmag2}).
The electric theory is an asymptotically free theory
and it is possible to take a decoupling limit $l_s\ra 0$
with the dynamical scale kept fixed by tuning the string coupling $g_s$
in a controlled way.
Here, $l_s$ is the string length, and all the stringy massive 
fields will be decoupled and higher dimensional terms will vanish
in this decoupling limit.

However, the magnetic theory is more tricky.
Since the negative mass squared for the tachyon field $t$ in
magnetic theory (I) is proportional to $1/l_s^2$,
it is not possible to take a smooth decoupling limit $l_s\ra 0$ that
pushes the string scale to infinity.
Magnetic theory (II) does not have a tachyonic mode. However, since
this theory is asymptotically non-free, it is again not clear how to take
the decoupling limit in a controlled way.
In this paper, we will not try to discuss these issues.
We treat this magnetic theory as a low-energy effective
theory with a cutoff scale around $1/l_s$, above which the theory
becomes the (possibly strongly coupled) string theory described
in \S \ref{O3D3bar}, and focus on the low-energy behavior
such as the vacuum structure and massless spectrum.
Since the coupling will become strong around the dynamical scale of
the magnetic theory, we set the cutoff scale $1/l_s$ to be around the
dynamical scale. In principle, it may be possible to make $l_s$ smaller,
because the decoupling limit $l_s\ra 0$ in the electric description
is well defined and the S-duality in string theory is believed
to be an exact duality. However, in any case, the
magnetic theory will become strongly coupled above the
dynamical scale and the perturbative analysis breaks down.
Therefore, we can only trust this magnetic description at low
energies.
Since this magnetic theory is asymptotically non-free, we expect that
the magnetic description will be better than the electric description at
low energies.

\subsection{Quantum corrections to the mass terms}
\label{oneloopmass}

As already mentioned in \S \ref{sec2}, quantum corrections
to the mass terms of the scalar fields are non-vanishing in both the
electric and magnetic theories, because the supersymmetry is completely
broken.
In our setup, the string scale $1/l_s$ plays the role of a natural
cutoff scale and the mass squared for the scalar fields turns out to be
finite and proportional to $1/l_s^2$. The open string one-loop calculation
was performed in Ref.~\citen{Uranga:1999ib},
and it was shown that the mass squared
for $\Phi^I$ in the electric theory is positive and that for $\phi^I$ in
magnetic theory (II) is negative:
\begin{eqnarray}
 m_{\Phi}^2= +C g_sl_s^{-2} \ ,~~ m_{\phi}^2= -C'g_sl_s^{-2} \ ,
\label{msq}
\end{eqnarray}
where $g_s$ is the string coupling, and
$C$ and $C'$ are positive numerical constants.

Note that the eigenvalues of the scalar fields correspond
to the positions of the \AD3-branes in the 6-dimensional
space transverse to the O3-plane.
The result (\ref{msq}) suggests that the \AD3-branes
are attractive and repulsive to the \Op3 and \Omh3-planes, respectively,
when the \AD3-branes are close to the O3-plane.
It was also shown in Ref.~\citen{Uranga:1999ib}
that this behavior is the same even when
\AD3-branes are far from the O3-plane,
at least in the open string one-loop calculation.

However, this result can be trusted only when the coupling
is small in each description.
Since the electric and magnetic theories
are asymptotically free and non-free, respectively,
we expect that the system will be better described at high and low
energies by the electric and magnetic theories, respectively.
Therefore, when \AD3-branes are placed near the origin of the
$x^{4\sim 9}$-plane, where the O3-plane is located,
we expect that the magnetic theory will be the better description and
then the \AD3-branes are repulsed from the origin.
On the other hand, when the \AD3-branes are far from the origin
of the $x^{4\sim 9}$-plane, the electric theory is the better
description and the \AD3-branes are attracted to the origin.
The fact that the \AD3-branes are repulsed from the \Omh3-plane
in the magnetic description implies that the scalar field $\phi^I$
in magnetic theory (II) will develop non-zero vacuum expectation
values in the magnetic description, which will break the $SO(6)$ symmetry.
In \S\ref{ConfDSBn=2} and \S\ref{ConfDSBn>2},  
we argue that this is how symmetry breaking caused by (\ref{TrQQ}) in
the electric description is realized in the magnetic description.

Note that $g_s$ is proportional to $\lambda/n$,
where $\lambda=g_{\rm YM}^2 n$ is the 't Hooft coupling,
and hence, the masses (\ref{msq}) are at subleading order
in the $1/n$ expansion.
This is consistent with the fact that these theories are equivalent
to the $\cN=4$ SYM at the leading order
of the $1/n$ expansion as discussed in \S \ref{first}.
However, if we take the decoupling limit $l_s\ra 0$
with fixed $n$, which only makes sense for
the electric theory as discussed in \S \ref{decoupling},
the scalar field $\Phi^I$ will decouple from the low-energy
physics.

\section{Confinement and dynamical symmetry breaking}
\label{sec4}

\subsection{Monopole condensation and confinement for $n=1$}

Let us first consider the $n=1$ case as a warm-up example.
As explained in \S \ref{eleth},
the electric theory flows to the $SU(2)$ pure Yang-Mills theory
(with gauge singlet massless fermions $Q^i$) at low energy.
On the other hand, magnetic theory (I) in Table \ref{mag1} is an
$SO(2)\simeq U(1)$ gauge theory for $n=1$.
After condensation of the tachyon field $t$, the $U(1)$ gauge symmetry
is broken, and all the fields, except for the
neutral components of the fermions $q^i$, will become massive.
Since the magnetic gauge group is completely broken, the electric theory
is expected to be confining. This is a manifestation of the dual
Meissner effect proposed in the 1970's.\cite{Nambu:1974zg,tHooft,Mandelstam:1974pi} 
In fact, the tachyon field $t$ is a magnetic monopole in the sense that
it is charged under the magnetic $U(1)$ gauge symmetry.

Note that the magnetic theory is an Abelian gauge theory,
although the electric theory is an $SU(2)$ gauge theory.
As discussed in \S \ref{first}, the $SU(2)$ gauge symmetry
will be broken to a $U(1)$ subgroup when we consider the configuration
with (\ref{Phi1}). The magnetic $U(1)$ gauge group is the
electric-magnetic dual of this unbroken $U(1)$ part of the $SU(2)$
gauge group in the electric theory. Because the scalar fields $\Phi^I$
acquire mass (\ref{msq}) via the quantum effect, the vacuum expectation
value of the scalar fields will roll down the potential to
the origin $v\ra 0$ and the $SU(2)$ gauge symmetry will eventually be
restored in the electric theory (in the perturbative picture).
However, the magnetic $U(1)$ gauge symmetry will not be enhanced
in the $v\ra 0$ limit in (\ref{magphi1}), since there is no
extra massless gauge field in the spectrum. Instead, a scalar field $t$
becomes tachyonic and causes the spontaneous breaking of the magnetic
$U(1)$ gauge symmetry when $v$ is small.

The situation is similar to the mass-deformed $\cN=2$ $SU(2)$ SYM
studied in Ref.~\citen{Seiberg:1994rs}.\footnote{
It may be useful to recall that this system can be
described in terms of the $SO$-type Seiberg duality
studied in Refs.~\citen{Seiberg:1994pq} and \citen{Intriligator:1995id}.
Note that the $\cN=2$ $SU(2)$
SYM can be regarded as an $\cN=1$ $SO(3)$ SQCD with one flavor of
chiral multiplet that belongs to the vector (=~adjoint)
representation of the $SO(3)$ gauge group. The magnetic dual of this
system is an $\cN=1$ $SO(2)(=SO(N_f-N_c+4))$ SQCD with one flavor of
magnetic monopole and a gauge-invariant meson field that couple through
a superpotential.}
The $\cN=2$ $SU(2)$ SYM can be regarded as the $\cN=1$
$SU(2)$ gauge theory with one chiral superfield $\Phi$
that belongs to the adjoint representation of the $SU(2)$ gauge group.
When the scalar component of $\Phi$ is at the generic point of the flat
direction of the potential, the $SU(2)$ gauge group is broken to its
$U(1)$ subgroup. If we deform the system by
adding a mass term for the chiral superfield $\Phi$, the flat direction
will be lifted and one may regard the system as the $\cN=1$ $SU(2)$ SYM
at low energies, which is believed to be a confining theory.
Since this system is an asymptotically free
theory, it is better to move to the magnetic description.
The magnetic theory is an $\cN=2$ $U(1)$ gauge theory
with a hypermultiplet charged under the magnetic $U(1)$ gauge
symmetry, which we call the magnetic monopole. 
The magnetic $U(1)$ gauge symmetry is the electric-magnetic dual
of the unbroken $U(1)$ subgroup of the $SU(2)$ gauge symmetry in the
electric description at the generic point of the vacuum moduli space.
This magnetic $U(1)$ symmetry will never become enhanced even at the
singularity of the vacuum moduli space.
The mass deformation for the chiral superfield $\Phi$ in the electric
description corresponds to adding a term in the superpotential of the
magnetic description that makes monopoles tachyonic and causes
the spontaneous breaking of the magnetic $U(1)$ gauge symmetry
via monopole condensation. This monopole condensation
is considered to be responsible for the confinement of the $\cN=1$
$SU(2)$ gauge theory in the electric description.\cite{Seiberg:1994rs}

\subsection{Flux tubes}
\label{fluxtube}

Our string theory setup is also useful for understanding the
properties of flux tubes.\footnote{
This subsection can also be skipped if readers want to avoid the
discussion using string theory.
}
Let us consider the potential between a pair consisting of
a heavy external quark
and antiquark\footnote{Here, a particle that belongs to the fundamental
representation of the $USp(2n)$ group is called a ``quark''.
Since the fundamental representation and the
antifundamental representation are equivalent for the $USp(2n)$ group,
there is actually no distinction between a quark and an antiquark.}
in the electric theory, which is obtained as the gauge theory realized in
the system with an \Op3-plane and $n$ \AD3-branes.
A heavy quark in the fundamental representation of the $USp(2n)$ gauge
group can be introduced by adding a fundamental string
stretched along the $x^9$ direction with one end attached
on the \AD3-branes. The end point of the fundamental string
behaves as an electrically charged point-like particle on the
\AD3-brane world-volume, which is interpreted as a ``quark''.
To introduce an ``antiquark'', we put another fundamental string
ending on one of the \AD3-branes.
We are interested in the behavior of the gauge flux when
the quark and antiquark are placed in fixed positions
separated by a large distance.\footnote{One way to fix the position
of the end point of the string may be to add
a D5-brane extended in the $x^0$ and $x^{4\sim 8}$ directions,
and consider a fundamental string stretched between the
\AD3-brane and the D5-brane. The following discussion does not
depend on how the end points of the strings are fixed.
}
The energy carried by the gauge flux is interpreted as the
potential energy of the quark-antiquark pair.
In the perturbative picture of the electric theory,
we observe only Coulomb-like potential as usual in the weakly
coupled gauge theory.
However, since the electric theory is an asymptotically free theory,
we can only trust this perturbative picture at high energies.

To determine the behavior at a low energy (large distance),
it is better to
move to the dual magnetic description.
Here, we consider the $n=1$ case for simplicity.
As explained in \S \ref{O3D3bar}, the magnetic description with
$n=1$ is given by an \Omh3-plane without additional \AD3-branes.
The quark (fundamental string) in the electric description corresponds
to a D-string in the magnetic description. 
Recall that the \Omh3-plane is understood as an \Om3-plane with a
spherical D5-brane that carries the magnetic flux
(\ref{magflux2}). Then, the argument in \S \ref{O3S} suggests that
D-strings cannot end on the \Omh3-plane.
This is again consistent with the confinement in the electric theory.
Because D-strings cannot end on the \Omh3-plane,
the D-strings corresponding to the quark and antiquark
should be connected by extending a D-string between them.
This D-string stretched between the quark and antiquark corresponds to
a color flux tube in the electric description.
Therefore, the potential of the quark and antiquark will be linear with
respect to the distance between them, because the energy carried by the
D-string is proportional to its length.
In other words, the Wilson loop in the electric description
('t Hooft loop in the magnetic description) exhibits the area law
behavior.
Note that the D-string tension is proportional to
$1/l_s^2$, where $l_s$ is the string length.
However, this does not necessarily mean that the tension is much higher
than the dynamical scale of the gauge theory. 
As mentioned in \S \ref{decoupling}, we do not try to take an
$l_s\ra 0$ limit in the magnetic theory but keep $l_s$ finite, because
the theory is asymptotically non-free.

There is, however, one subtlety here. The open strings
stretched between the D-string and a half \AD3-brane stuck at the
\Om3-plane have a tachyonic mode.
A similar tachyonic mode exists in the case of a parallel D1-D3 system
(without an O3-plane), in which case 
the condensation of the tachyonic mode corresponds to dissolving the
D-string in the D3-brane world-volume as a uniformly distributed
magnetic flux. In our case with the \Omh3-plane, it is not possible
to replace the D-string with the magnetic flux, since there is
no gauge field on the \Omh3-plane. Although we are not able
to prove this explicitly, we suppose that the stable configuration
is an analogue of the D1-D3-D5 bound state considered in
Ref.~\citen{Polchinski:2000uf} and
that it behaves as a string-like object
with finite tension. This interpretation is more plausible
for the $n>1$ cases discussed in the following subsections.

Next, we consider two quarks placed at the same point
in the electric description.
Since the gauge group in the electric description is $USp(2n)$,
and all the fields are in a rank-2 tensor representation
of the gauge group, a color charge in the fundamental representation
cannot be screened, and in fact we have observed the
linear potential between a quark and an antiquark above.
However, if we consider two quarks, the color charge will be screened
by the gauge field and/or the matter fields.
Note that the gauge group $USp(2n)$ has a center
$\Z_2$ and that all the matter fields in the electric theory (Table \ref{ele})
are invariant under this $\Z_2$.
We expect that the Wilson loop for the quark in the representation $R$
of the gauge group will exhibit the area or perimeter law
when $R$ is non-trivial or trivial with respect to this $\Z_2$,
respectively. The question is whether we can understand this $\Z_2$
property using string theory.

In fact, the answer is yes. As we have seen, the color flux tube between
a quark and an antiquark is given by the D-string in the magnetic
description. It is known that the D-string stretched along 
the \Omh3-plane is a $\Z_2$-charged object. Namely, one D-string
is stable but two D-strings are unstable.
In fact, this D-string is related by T-duality to a non-BPS D7-brane
in type I string theory, whose charge is classified by $KO(\R^2)\simeq
\Z_2$.\cite{Witten:1998cd}
To see this more explicitly, following Ref.~\citen{Bergman:2000tm},
let us consider $k$ D-strings stretched along the \Omh3-plane.
The mirror image of the D-strings under the orientifold action
will be D-strings with the opposite orientation (\AD1-branes).
It is known that the gauge theory realized on the D-string
world-sheet is a $U(k)$ gauge theory with a tachyon field
in the antisymmetric tensor representation.\footnote{
A useful table can be found
in Ref.~\citen{Bergman:2000tm} and also in Appendix A of
Ref.~\citen{Imoto:2009bf}.}
Since the antisymmetric tensor representation of $U(k)$
does not exist for $k=1$,
there is no tachyon field and the D-string is stable for the $k=1$ case,
but the $k=2$ case is tachyonic and unstable.
In general, for odd $k$, the rank of the tachyon field is at most $(k-1)$,
and hence $(k-1)$ D-strings can be annihilated but one D-string
will remain stable.
For even $k$, all the D-strings are annihilated
after the condensation of the tachyon field on the D-strings,
and the Wilson loop (in the electric description) will no longer exhibit
the area law.

Note that magnetic theory (II) is obtained by Higgsing
from magnetic theory (I), which is a $U(1)$ gauge theory
for $n=1$. Since the vortices in the Abelian Higgs model
are classified by $\pi_1(U(1))\simeq\Z$, one might think
that the flux tube should be classified by $\Z$ rather than $\Z_2$.
However, as argued in Ref.~\citen{Witten:1998cd},
the topological classification of the D-branes
in string theory is given by K-theory.
In our case, the K-theory group classifying the flux tubes is
$KO(\R^2)\simeq\Z_2$,
which is consistent with what we expect in the electric theory.
This fact seems to suggest that we should take into account the
creation and annihilation of D3-\AD3 pairs to obtain
the correct topological classification of the flux
tubes.\footnote{A similar observation has been found in the holographic
dual of the $SO(N_c)$ QCD in Ref.~\citen{Imoto:2009bf}.}

\subsection{A toy model}
\label{toy}

As we have seen around (\ref{TrQQ}), the global $SO(6)$ symmetry is
expected to be dynamically broken to the $SO(4)$ subgroup for the $n>1$
cases.
The basic idea to understand the dynamical symmetry breaking using the
S-duality was already explained in \S \ref{oneloopmass}.
The perturbative calculation in the magnetic theory suggests that the
scalar field $\phi^I$ is unstable around the origin and will develop a
non-zero vacuum expectation value that causes the breaking of the
$SO(6)$ symmetry. On the other hand, perturbative analysis of the
electric theory shows that the \AD3-branes are attracted to the
\Op3-plane, which suggests that the potential for the scalar field
$\phi^I$ will increase for large $\phi^I$, and we expect that
a minimum of the potential exists somewhere in between.
It is, however, not easy to show which configuration minimizes the energy,
since we do not know the precise form of the
potential for the scalar field $\phi^I$. The minimum of the potential
is expected to exist in the region where neither electric nor magnetic
descriptions are weakly coupled.
Therefore, instead of trying to find the precise potential,
we consider a toy model that captures qualitative features of the
potential described above and argue that the symmetry breaking
$SO(6)\simeq SU(4)\ra SO(4)$ expected from (\ref{TrQQ}) can occur
naturally.

The model we consider is based on magnetic theory (II), whose field
content is as listed in Table \ref{mag2}. We consider the following
potential for the scalar field $\phi^I$:
\begin{eqnarray}
 V(\phi^I)= -\frac{\mu^2}{2}\tr(\phi^I\phi^I)
-\frac{g}{4}\tr\left([\phi^I,\phi^J]^2\right)
+\frac{\lambda}{2}\tr\left((\phi^I\phi^I)^2\right)\ ,
\label{Vphi}
\end{eqnarray}
where $\mu^2$, $g$ and $\lambda$ are all positive constants,
and the repeated indices are summed over.
Here, $\phi^I$ is an antisymmetric pure imaginary $(2n-1)\times (2n-1)$
matrix valued scalar field
and $I(=1,2,\cdots,6)$ is the vector index for the global $SO(6)$ symmetry.
The signs of the couplings are chosen to realize the
qualitative features of magnetic theory (II).
The first term is the tachyonic mass term, which is suggested to be
generated via the quantum effect as argued in \S \ref{oneloopmass}.
The second term is the same as the tree-level potential in
(\ref{Lmag2}).
The third term is added to stabilize the potential, which
reflects the fact that the \AD3-branes are attracted to the \Op3-plane
at a large distance as explained above.
If we seriously calculate the quantum corrections to the potential, 
many other terms will be generated. In the following, we simply
discard the other possible terms and analyze the potential (\ref{Vphi}),
hoping that the qualitative properties of the magnetic theory
are correctly captured.

We are interested in static solutions that minimize the potential
energy. The equation of motion obtained by differentiating the potential
(\ref{Vphi}) is
\begin{eqnarray}
-\mu^2\phi^I
-g [\phi^J,[\phi^I,\phi^J]]
+\lambda\left( \phi^I(\phi^J\phi^J)
+ (\phi^J\phi^J)\phi^I
\right)=0\ .
\label{eom}
\end{eqnarray}
The potential energy for the configurations satisfying this equation of
motion is
\begin{eqnarray}
 V(\phi^I)=-\frac{\mu^2}{4}\tr(\phi^I\phi^I)\ .
\label{pot}
\end{eqnarray}

\subsection{Confinement and dynamical symmetry breaking for $n=2$}
\label{ConfDSBn=2}

Let us first analyze the case with $n=2$.
In this case, the scalar fields $\phi^I$ are
$3\times 3$ antisymmetric matrices, which can be expanded as
\begin{eqnarray}
 \phi^I=\wt A^I_i J^i\ ,
\label{phiAJ}
\end{eqnarray}
where $\wt A^I_{i}\in \R$ $(i=1,2,3)$ and
\begin{eqnarray}
J^1=
i\left(
\begin{array}{ccc}
0&0&0\\
0&0&-1\\
0&1&0
\end{array}
\right)\ ,
~~
J^2=
i\left(
\begin{array}{ccc}
0&0&1\\
0&0&0\\
-1&0&0
\end{array}
\right)\ ,
~~
J^3=
i\left(
\begin{array}{ccc}
0&-1&0\\
1&0&0\\
0&0&0
\end{array}
\right)\ .
\label{J1}
\end{eqnarray}
Note that these $J^i$ satisfy
\begin{eqnarray}
 [J^i,J^j]=i\epsilon^{ijk}J^k\ ,
\label{su2}
\end{eqnarray}
and form the spin-1 representation of the $su(2)$ algebra.
The $SO(3)$ gauge symmetry acting on the scalar field $\phi^I$
is converted to the $SO(3)$ rotation acting on index $i$ of
$\wt A^I_i$.

Using the useful relation
\begin{eqnarray}
 J^i\{J^j,J^k\}+ \{J^j,J^k\}J^i=2\delta^{jk}J^i
+\delta^{ij}J^k+\delta^{ik}J^j\ ,
\end{eqnarray}
the equation of motion (\ref{eom}) yields
\begin{eqnarray}
 -\mu^2\wt A_i^I+(\lambda-g)\wt A^I_l\wt A^J_l\wt A^J_i
+(\lambda+g)\wt A^J_l\wt A^J_l\wt A^I_i=0\ .
\label{eomA}
\end{eqnarray}
Using the $SO(6)$ symmetry, we can 
always set $\wt A^m_i=0$ for $m=4,5,6$. In this case,
(\ref{eomA}) implies
\begin{eqnarray}
X\left(-\mu^2 1_3+(\lambda-g)X+
(\lambda+g)(\tr X)1_3\right) =0\ ,
\label{eomX}
\end{eqnarray}
where we have defined $X=(X^{ij})\equiv (\wt A^i_k\wt A^j_k)$,
$(i,j,k=1,2,3)$.
Since $X$ is a real symmetric $3\times 3$ matrix by definition,
it can be diagonalized by using the $SO(3)$ symmetry.
Then, from (\ref{eomX}) we see that all the non-zero eigenvalues of $X$
should be the same for $\lambda\ne g$.
Using this fact, it is not difficult to show that the general
non-trivial solutions of (\ref{eom}) can be written as
\begin{eqnarray}
&\mbox{(A)}&~~ \phi^1=a J^1\ ,~~\phi^{2\sim 6}=0\ ,
\label{sol(A)}\\
&\mbox{(B)}&~~ \phi^1=a J^1\ ,~~ \phi^2=a J^2\ ,~~\phi^{3\sim 6}=0\ ,
\\
&\mbox{(C)}&~~ \phi^1=a J^1\ ,~~ \phi^2=a J^2\ ,~~ \phi^3=a J^3
\ ,~~\phi^{4\sim 6}=0\ ,
\label{sol(C)}
\end{eqnarray}
up to the $SO(3)\times SO(6)$ symmetry. The value of the coefficient $a$
and the potential energy for these solutions are
\begin{eqnarray}
&\mbox{(A)}&~~ a^2= \frac{\mu^2}{2\lambda}\ ,~~~
V(\phi^I)=- \frac{\mu^4}{4\lambda}\ ,
\\
&\mbox{(B)}&~~ a^2= \frac{\mu^2}{g+3\lambda}\ ,~~~
V(\phi^I)=- \frac{\mu^4}{g+3\lambda}\ ,
\\
&\mbox{(C)}&~~ a^2= \frac{\mu^2}{2(g+2\lambda)}\ ,~~~
V(\phi^I)=- \frac{3\mu^4}{4(g+2\lambda)}\ .
\label{(C)}
\end{eqnarray}

Solutions (A) and (C) are the 
lowest energy configurations for $\lambda<g$ and $\lambda>g$,
respectively.
Solution (A) corresponds to the configuration with
a half \AD3-brane stuck at the \Om3-plane and
a \AD3-brane separated from it.
Solution (C) is a fuzzy sphere configuration
and it is interpreted as a spherical D5-brane
blown up to a finite size with the \AD3-brane absorbed
as the magnetic flux on it.
Note that this fuzzy sphere configuration is very similar to the
spherical D5-brane in the Polchinski-Strassler model
\cite{Polchinski:2000uf} obtained via the Myers effect
\cite{Myers:1999ps}. Unlike the Myers effect,
we have not added additional RR-flux to inflate the spherical D-brane.
Here, the sphere has been blown up because of the tachyonic mass
term in the potential (\ref{Vphi}) that represents the repulsive force
between the \Om3-plane and \AD3-branes.

Since we are working with a toy model, it is not possible to show which
configuration is realized in our brane configuration discussed
in \S \ref{O3D3bar}.
In the following, we assume $\lambda>g$, for which solution (C) is
realized, and show that this configuration gives a qualitatively
consistent picture with the confinement and dynamical symmetry
breaking expected in the electric theory.

In solution (C), the global symmetry $SO(6)$ is broken
to $SO(3)_{1\sim 3}\times SO(3)_{4\sim 6}$.
The first factor $SO(3)_{1\sim 3}$
is the rotation of $\phi^{1\sim 3}$ compensated
by the action of the $SO(3)$ gauge group to keep the vacuum expectation
value (\ref{sol(C)}) fixed.
The second factor $SO(3)_{4\sim 6}$ is simply
the rotation of $\phi^{4\sim 6}$.
Note that $SO(3)\times SO(3)$ is equivalent to
$SO(4)\simeq (SU(2)\times SU(2))/\Z_2$ (at least locally),
which is the unbroken symmetry expected from (\ref{TrQQ}).
In fact, one can check that the generators of the $SO(3)\times SO(3)$
subgroup in the spinor representation correspond to those of the $SO(4)$
subgroup of $SU(4)$. (See Appendix \ref{Sigma}.)

Up to now we have been a little sloppy in our notation of the $SO(N)$
group, because we have not distinguished it from its universal covering.
Let us elaborate on this issue here. Since the $SO(6)$ symmetry acts on
the fermion as a spinor representation, the global symmetry is actually
$SU(4)$ rather than $SO(6)$.
The subgroup that keeps the vacuum expectation value (\ref{TrQQ})
invariant in the electric theory is the $SO(4)$ subgroup,
which is the real part of $SU(4)$.
On the other hand, if we do not take into account the fields
that belong to the spinor representation,
the unbroken symmetry in the magnetic theory is the
$SO(3)\times SO(3)$ subgroup of $SO(6)$.
Note that the spinor representation of $SO(6)$
transforms as the bi-spinor representation of the $SO(3)\times SO(3)$
subgroup, which is equivalent to the bi-fundamental representation of
its universal covering $SU(2)\times SU(2)$. Because
the diagonal $\Z_2$ element acts trivially on the bi-fundamental
representation of $SU(2)\times SU(2)$, the unbroken subgroup is
 $(SU(2)\times SU(2))/\Z_2\simeq SO(4)$ as expected.

In the fuzzy sphere configuration (\ref{sol(C)}), the $SO(3)$ gauge
symmetry is completely Higgsed. This is again consistent with the
confinement in the electric theory. Therefore, the dynamical symmetry
breaking and confinement are both caused by the vacuum expectation value
of the scalar field in the magnetic description.\footnote{
A similar phenomenon was found in a recent proposal
to analyze QCD using Seiberg duality with soft SUSY breaking
deformations.\cite{Kitano:2011zk}}

These phenomena can be understood geometrically using the interpretation
of the fuzzy sphere configuration as the spherical D5-brane with
magnetic flux.\cite{Myers:1999ps}$^,$\footnote{The D5-brane world-volume
description is a good description when the size of the sphere is large
compared with the string length scale. In our case, the size may not be
large enough to justify the validity of the D5-brane world-volume description.
However, we think this D5-brane viewpoint is still worth mentioning,
because this geometric picture is useful for obtaining an intuitive
understanding. }
As explained in \S \ref{O3D3bar}, the \Omh3-plane
can be thought of as an \Om3-plane with a spherical D5-brane.
We expect that \AD3-branes will be absorbed in the spherical D5-brane
and make it blow up to a finite size because of the repulsive force
between the \Om3-plane and \AD3-branes.
Recall that in our brane configuration, the $SO(6)$ symmetry corresponds
to the rotation of the $x^{4\sim 9}$-plane.
If the spherical D5-brane is embedded in the $x^{4\sim 6}$-plane,
the rotational symmetry is broken to
$SO(3)_{1\sim 3}\times SO(3)_{4\sim 6}$, where
$SO(3)_{1\sim 3}$ and $SO(3)_{4\sim 6}$ correspond to the rotation
of $x^{4\sim 6}$ and $x^{7\sim 9}$-planes, respectively.
Since D-strings cannot end on the D5-brane, the linear potential for
the quark-antiquark pair is obtained in the same way as the $n=1$
case discussed in \S \ref{fluxtube}.

Let us next consider the fluctuations around the fuzzy sphere
solution (\ref{sol(C)}):
\begin{eqnarray}
\begin{array}{ll}
 \phi^i(x)=a J^i +\delta \phi^i(x)\ ,& (i=1,2,3)\\
 \phi^m(x)=\delta \phi^m(x)\ ,& (m=4,5,6)
\label{fluc1}
\end{array}
\end{eqnarray}
where $a$ is as given in (\ref{(C)}).
Again, the fluctuations $\delta\phi^I(x)$ can be expanded as
\begin{eqnarray}
 \delta\phi^I(x)= A^I_i(x)J^i\ ,
\label{deltaphiAJ}
\end{eqnarray}
where $A^I_i(x)\in\R$.
Inserting this configuration into the potential (\ref{Vphi}), we obtain
\begin{eqnarray}
 V(\phi^I)&=&-\frac{3\mu^4}{4(g+2\lambda)}
+\frac{\mu^2}{2(g+2\lambda)}
\left(
(\lambda-g)
(A^j_iA^j_i+A^j_iA^i_j)
+2(\lambda+g)(A^i_i)^2
\right)\nn\\
&&+{\cal O}(A^3)\ ,
\label{flucA}
\end{eqnarray}
where $i,j=1,2,3$.
Here, we have used the relation
\begin{eqnarray}
 \tr(J^iJ^jJ^kJ^l)=\delta^{ij}\delta^{kl}+\delta^{il}\delta^{jk}\ .
\end{eqnarray}
Note that the 9 components $A^m_i$ ($m=4,5,6$; $i=1,2,3$)
and the antisymmetric part of the real $3\times 3$ matrix $A\equiv(A^i_j)$
($i,j=1,2,3$) do not have mass terms. The former correspond to the
Nambu-Goldstone modes associated with the symmetry breaking
$SO(6)\ra SO(3)\times SO(3)$.
The latter correspond to the direction of the gauge rotation (would-be
Nambu-Goldstone modes),
which will be absorbed in the $SO(3)$ gauge field to make them massive.
It is useful to decompose $A=(A^i_j)$ as
\begin{eqnarray}
 A=\xi 1_3 +\eta+\chi\ ,
\end{eqnarray}
where $\eta$ and $\chi$ are real $3\times 3$ traceless symmetric and
antisymmetric matrices, respectively, and $\xi\in\R$ is the trace part
of matrix $A$.
Then the potential (\ref{flucA}) becomes
\begin{eqnarray}
V(\phi^I)
=-\frac{3\mu^4}{4(g+2\lambda)}
+\frac{\mu^2}{2(g+2\lambda)}
\left(
12(g+2\lambda)\xi^2
+2(\lambda-g)
\tr(\eta^2)
\right)
+{\cal O}(A^3)\ .
\label{j=1pot}
\end{eqnarray}
The trace part $\xi$, which corresponds to the fluctuation of
the radius of the fuzzy sphere, is always massive.
The traceless symmetric part $\eta$ is also massive when
$\lambda>g$. There is no tachyonic mode in this case.

\subsection{Confinement and dynamical symmetry breaking for $n>2$}
\label{ConfDSBn>2}

Although the analysis for the $n>2$ case is more complicated,
the basic story is the same as that for the $n=2$ case.
We will show that if $\lambda$ is large enough,
the fuzzy sphere configuration
\begin{eqnarray}
\begin{array}{ll}
\phi^i=a J_{(n-1)}^i\ ,& (i=1,2,3)\\
\phi^m=0\ ,& (m=4,5,6)
\label{fuzzy}
\end{array}
\end{eqnarray}
minimizes the potential energy.
Here, $J_{(n-1)}^i$ $(i=1,2,3)$ are the generators of the $su(2)$ algebra
(\ref{su2}) in the spin-$(n-1)$ representation, and the coefficient $a$
will be determined in (\ref{a}).
This configuration corresponds to a spherical D5-brane with $-(n-1/2)$
units of magnetic flux.
Then, the dynamical symmetry breaking and confinement are explained in
exactly the same way as in the previous subsection.
Note that the scalar fields $\phi^I$ are $(2n-1)\times (2n-1)$
antisymmetric matrices and the spin-$(n-1)$ representation is the
$(2n-1)$-dimensional irreducible representation.
Schur's lemma implies that the $SO(2n-1)$ gauge symmetry
is completely Higgsed for the fuzzy sphere configuration
(\ref{fuzzy}), suggesting that 
the electric description is confined via the dual Meissner effect.

In the following, we start with examining some of the solutions
of the equation of motion (\ref{eom}), including
the configurations with isolated \AD3-branes, multiple fuzzy spheres
and their combinations. We will show that the fuzzy sphere solution
(\ref{fuzzy}) is the least-energy configuration among the
solutions we find, and stable against small fluctuations when $\lambda$
is large enough, although we have not succeeded in confirming that it is
the global minimum of the potential.
We also identify the Nambu-Goldstone modes as well as the would-be
Nambu-Goldstone modes eaten by the gauge field.

\subsubsection{Isolated \AD3-branes}
\label{iso}

First, we consider a solution of the equation of motion (\ref{eom})
that corresponds to the configuration
with $(n-1)$ \AD3-branes put away from the origin in the $x^4$-direction:
\begin{eqnarray}
\phi^1=v
\left(
\begin{array}{ccccc}
\sigma_2\\
&\sigma_2\\
&&\ddots\\
&&&\sigma_2\\
&&&&0
\end{array}
\right)\ ,~~~\phi^{2\sim 6}=0\ .
\end{eqnarray}
Inserting this into the equation of motion (\ref{eom}),
we obtain
\begin{eqnarray}
v^2=\frac{\mu^2}{2\lambda}\ .
\end{eqnarray}
The energy for this solution calculated with (\ref{pot}) is
\begin{eqnarray}
V(\phi^I)=-(n-1)\frac{\mu^4}{4\lambda}\ .
\label{isoD3}
\end{eqnarray}

\subsubsection{One fuzzy sphere}
\label{onefuzz}

Since the generators of the $su(2)$ algebra in the spin-$j$ representation
$J^i_{(j)}$ ($i=1,2,3$) satisfy the relation
\begin{eqnarray}
 J_{(j)}^iJ_{(j)}^i=j(j+1)1_{2j+1}\ ,
\end{eqnarray}
the fuzzy sphere configuration (\ref{fuzzy}) satisfies
\begin{eqnarray}
 \phi^J\phi^J=a^2 n(n-1) 1_{2n-1}\ .
\label{c2}
\end{eqnarray}
Inserting (\ref{fuzzy}) and (\ref{c2}) into (\ref{eom}),
we obtain
\begin{eqnarray}
\left(-\mu^2+2ga^2+2\lambda a^2 n(n-1)\right)\phi^I=0\ ,
\end{eqnarray}
which shows that this configuration is a solution of the
equation of motion (\ref{eom}) if
\begin{eqnarray}
 a^2=\frac{\mu^2}{2(g+\lambda n(n-1))}\ .
\label{a}
\end{eqnarray}
The energy carried by this configuration is
\begin{eqnarray}
V(\phi^I)=
-\frac{\mu^4 n(n-1)(2n-1)}{8(g+\lambda n(n-1))}\ .
\label{oneS2}
\end{eqnarray}
Comparing this with (\ref{isoD3}), we see that
the fuzzy sphere configuration is favored for
\begin{eqnarray}
 \lambda n> 2g\ .
\label{lng}
\end{eqnarray}

\subsubsection{Multiple fuzzy spheres}

As a generalization of the fuzzy sphere solution (\ref{fuzzy}),
consider a configuration with $k$ fuzzy spheres:
\begin{eqnarray}
\phi^i=
\left(
\begin{array}{ccccc}
a_1J_{(j_1)}^i\\
&a_2J_{(j_2)}^i\\
&&\ddots\\
&&&a_kJ_{(j_k)}^i\\
\end{array}
\right)\ ,~~(i=1,2,3)\ ,~~~\phi^{4\sim 6}=0,
\end{eqnarray}
where $J_{(j_r)}^i$ are the generators of
the $su(2)$ algebra in the spin-$j_r$ representation.\footnote{
Here, the scalar fields $\phi^I$ are block-diagonalized
with respect to the irreducible representations of the $su(2)$ algebra
for convenience.
Since $\phi^I$ are originally pure imaginary antisymmetric matrices,
the representations of the $su(2)$ algebra with half odd integer spin
appear in complex conjugate pairs.}
Since $J_{(j_r)}^i$ are matrices of size $(2j_r+1)$,
the spins $\{j_r\}$ satisfy
\begin{eqnarray}
\sum_{r=1}^k(2j_r+1) = 2n-1\ .
\label{jn}
\end{eqnarray}
Since the right-hand side is an odd number, the number of spheres $k$
should be odd.
This configuration is a solution of the equation of motion (\ref{eom})
provided the coefficients $a_r$ satisfy
\begin{eqnarray}
 a_r^2=\frac{\mu^2}{2(g+\lambda j_r(j_r+1))}\ .
\end{eqnarray}

The energy (\ref{pot}) is then
\begin{eqnarray}
V(\phi^I)
=-\frac{\mu^4}{8}
\sum_{r=1}^k \frac{j_r(j_r+1)(2j_r+1)}{(g+\lambda j_r(j_r+1))}
=\sum_{r=1}^k (2j_r+1)\rho(j_r)
\ ,
\label{VkS2}
\end{eqnarray}
where
\begin{eqnarray}
\rho(j_r)=-\frac{\mu^4}{8}
\frac{j_r(j_r+1)}{(g+\lambda j_r(j_r+1))}
\end{eqnarray}
is the energy of the $r$th fuzzy sphere
per size of the matrix $(2j_r+1)$.
We can think of (\ref{VkS2}) as the summation of
the ``energy density'' $\rho(j_r)$ times the ``length'' $(2j_r+1)$.
The problem is to find the minimal energy configuration
when the ``total length'' is fixed by (\ref{jn}).
Since the ``energy density'' $\rho(j)$ is a
negative monotonically decreasing function
(the absolute value $|\rho(j)|$ is increasing)
with respect to $j$,
the minimal energy configuration in (\ref{VkS2}) is
clearly the $k=1$ case.
Therefore, one maximal-size sphere is always favored compared with
many smaller spheres.

\subsubsection{One fuzzy sphere and isolated \AD3-branes}
\label{combi}

Let us next consider
the combination of the configurations
considered in \S\ref{iso} and \S\ref{onefuzz}: 
\begin{eqnarray}
\phi^1=
\left(
\begin{array}{ccccc}
aJ_{(j)}^1\\
&v\sigma_2\\
&&\ddots\\
&&&v\sigma_2\\
\end{array}
\right)\ ,~~~
\phi^{2,3}=
\left(
\begin{array}{ccccc}
aJ_{(j)}^{2,3}\\
&0\\
&&\ddots\\
&&&0\\
\end{array}
\right)\ ,~~
\phi^{4\sim 6}=0\ ,
\nn\\
\end{eqnarray}
where
\begin{eqnarray}
 a^2=\frac{\mu^2}{2(g+\lambda j(j+1))}\ ,~~~v^2=\frac{\mu^2}{2\lambda}\ .
\end{eqnarray}
The number of $v\sigma_2$ in $\phi^1$ is
$(n-j-1)$.
The energy (\ref{pot}) is then estimated as
\begin{eqnarray}
V(\phi^I)=-(n-j-1)\frac{\mu^4}{4\lambda}
-\frac{\mu^4j(j+1)(2j+1)}{8(g+\lambda j(j+1))}\ .
\end{eqnarray}
We recover (\ref{isoD3}) and (\ref{oneS2})
with $j=0$ and $j=n-1$, respectively.

To estimate the value of $j$ that minimizes the energy, consider
the derivative of $V$ with respect to $j$
\begin{eqnarray}
\frac{\del V(\phi^I)}{\del j}
=\frac{\mu^4g(2g-(1+2j(j+1))\lambda)}{8\lambda(g+j(j+1)\lambda)^2}\ .
\label{delV}
\end{eqnarray}
Therefore, if
\begin{eqnarray}
\lambda > 2g\ ,
\end{eqnarray}
(\ref{delV}) is always negative and the sphere tends to expand to
the maximal size.
If
\begin{eqnarray}
\lambda < 2g\ , 
\end{eqnarray}
(\ref{delV}) is positive around $j=0$ and becomes negative for large
enough $j$, and hence the sphere will either shrink to zero size
or expand to the maximal size. As we have seen in \S \ref{onefuzz},
the fuzzy sphere with the maximal size is favored if (\ref{lng}) is
satisfied.

\subsubsection{Fluctuations around one fuzzy sphere}

We have seen that the fuzzy sphere solution (\ref{fuzzy})
is the minimal energy configuration among those considered
in \S\ref{iso}$\sim$\S\ref{combi} 
if $\lambda$ and $g$ satisfy (\ref{lng}). 
Let us next consider the fluctuations around the fuzzy sphere solution
(\ref{fuzzy}) and confirm that there are no tachyonic modes for
large enough $\lambda$.

Consider the fluctuations around the fuzzy sphere solution
 (\ref{fuzzy}), as we did in (\ref{fluc1}) for the $n=2$ case:
\begin{eqnarray}
\begin{array}{ll}
 \phi^i(x)=a J_{(n-1)}^i +\delta \phi^i(x)\ ,& (i=1,2,3)\\
 \phi^m(x)=\delta \phi^m(x)\ ,& (m=4,5,6)
\label{fluc}
\end{array}
\end{eqnarray}
where $a$ is given by (\ref{a}).
In this subsection, we set $j\equiv n-1$ and $J^i\equiv J_{(j)}^i=J_{(n-1)}^i$.

Inserting (\ref{fluc}) into the potential
(\ref{Vphi}), and using the relation (\ref{a}),
the terms quadratic in $\delta\phi^I$
are obtained as
\begin{eqnarray}
V|_{\cO(\delta\phi^2)}
&=&
-ga^2
\tr\left(\delta\phi^m\delta\phi^m
+\half[J^i,\delta\phi^m]^2\right)
\nn\\
&&
-\frac{ga^2}{2}
\tr\bigg(
\left(\delta\phi^i+i\epsilon^{ijk}
[J^{j},\delta\phi^{k}]\right)
\left(2\delta\phi^i-i\epsilon^{ij'k'}
[J^{j'},\delta\phi^{k'}]\right)
\bigg)
\nn\\
&&
+\frac{\lambda a^2}{2}\tr
\left(\{J^i,\delta\phi^i\}^2
\right)\ .
\label{Vphi2}
\end{eqnarray}
Here, the ranges of the indices are $i,j,k=1\sim 3$ and
$m=4\sim 6$.
See Appendix \ref{Junk} for the calculation.

\vspace{2ex}
\noindent {\bf Would-be Nambu-Goldstone modes}

Because the gauge symmetry is spontaneously broken
by the vacuum expectation value of the scalar fields,
$\dim (SO(2n-1))=(2n-1)(n-1)$ components of the fluctuation
in the scalar fields
will be absorbed into the gauge field. These components
(would-be Nambu-Goldstone modes)
correspond to the direction of the gauge rotation
\begin{eqnarray}
\delta\phi^I=[X,\VEV{\phi^I}]\ ,
\label{gauge}
\end{eqnarray}
or, more explicitly,
\begin{eqnarray}
\begin{array}{ll}
\delta\phi^i=a [X,J^i]\ ,& (i=1,2,3)\\
\delta\phi^m=0\ ,& (m=4,5,6)
\label{gauge2}
\end{array}
\end{eqnarray}
where $X\in so(2n-1)=so(2j+1)$ is a real antisymmetric matrix.
Then, it is easy to check that
\begin{eqnarray}
\{J^i,\delta\phi^i\}=0\ ,
\end{eqnarray}
and
\begin{eqnarray}
\delta\phi^i+i \epsilon^{ijk}[J^j,\delta\phi^k]=0\ .
\label{dphieigen}
\end{eqnarray}
Therefore, the potential (\ref{Vphi2}) is flat
along this direction as it should be.

\vspace{2ex}
\noindent {\bf Nambu-Goldstone modes}

Because of the symmetry breaking $SO(6)\ra SO(3)\times SO(3)$,
there are 9 components of the Nambu-Goldstone modes that remain massless.
The fluctuations corresponding to these Nambu-Goldstone modes are of the form
\begin{eqnarray}
\begin{array}{ll}
\delta\phi^i=0\ ,& (i=1,2,3)\\
\delta\phi^m=A^m_iJ^i\ ,& (m=4,5,6)
\label{NG}
\end{array}
\end{eqnarray}
where $(A^m_i)$ is a real $3\times 3$ matrix.
For this, we can show
\begin{eqnarray}
[J^i,[J^i, \delta\phi^m]]=2\delta\phi^m\ ,
\end{eqnarray}
which implies $V|_{\cO(\delta\phi^2)}=0$, and
these components are massless degrees of freedom
as expected.

\vspace{2ex}
\noindent {\bf General fluctuation}

As a generalization of (\ref{deltaphiAJ}), it is known that any 
$(2j+1)\times(2j+1)$ antisymmetric matrix $\delta\phi^I$
can be expanded as
\begin{eqnarray}
 \delta\phi^I
=\sum_{l:{\rm odd}}^{2j-1} \delta\phi^I_l\ ,
~~~~
\delta\phi^I_l=A^I_{i_1,i_2,\cdots,i_l}J^{i_1}J^{i_2}\cdots J^{i_l}\ ,
\label{Yexp}
\end{eqnarray}
where the coefficient $A^I_{i_1,i_2,\cdots,i_l}$ is a
real rank-$l$ traceless symmetric tensor with respect to the $SO(3)$
indices $i_1,i_2,\cdots,i_l$. This is equivalent to the expansion with
respect to the spherical harmonics for the fuzzy sphere.\footnote{See,
for example, Ref.~\citen{Dasgupta:2002hx} for related analysis in the
pp-wave matrix model and useful formulae for the fuzzy spherical
harmonics.}
Note that for each $I\in\{1,2,\cdots,6\}$, the traceless symmetric tensor
 $A^I_{i_1,i_2,\cdots,i_l}$ belongs to the spin-$l$ representation of
 $SO(3)$ acting on the lower indices $i_k$ ($k=1,2,\cdots,l$), and
the number of independent parameters is $(2l+1)$. By summing over
$l=1,3,5,\, \cdots,2j-1$, 
the total number of parameters is
\begin{eqnarray}
 \sum_{l:{\rm odd}}^{2j-1}(2l+1)
=  j(2j+1)\ ,
\end{eqnarray}
which agrees with that of $(2j+1)\times (2j+1)$
antisymmetric matrices.

First we consider $\delta\phi^I$ with $I=m=4,5,6$.
{}From the fact that $A^I_{i_1,\cdots,i_l}$ belongs to the spin-$l$
representation of $SO(3)$, we have
\begin{eqnarray}
[J^i,[J^i,\delta\phi_l^I]]
=l(l+1)\delta\phi_l^I\ .
\end{eqnarray}
Using this relation, the potential (\ref{Vphi2}) for $\delta\phi_l^m$
($m=4,5,6$) is obtained as
\begin{eqnarray}
 V|_{\cO(\delta\phi^2)}=\left(\half l(l+1)-1\right)ga^2
\tr\left(\delta\phi_l^m\delta\phi_l^m\right)\ .
\end{eqnarray}
Therefore, $l=1$ modes are massless and all the other modes with $l>1$ are
massive. The $l=1$ components correspond to the Nambu-Goldstone modes
considered in (\ref{NG}).

Next, we consider $\delta\phi^I$ with $I=i=1,2,3$. It is useful to
decompose $\delta\phi_l^i$ as
\begin{eqnarray}
 \delta\phi^i_l=\sum_{\epsilon=+1,0,-1} \delta\phi^i_{l(\epsilon)}\ ,
\end{eqnarray} 
following the branching rule for the tensor product of the spin-$l$ and
spin-$1$ representations of $SO(3)$: 
\begin{eqnarray}
\mbox{spin}~ l \otimes\mbox{spin}~1 =
\mbox{spin}~(l+1)\oplus\mbox{spin}~l\oplus\mbox{spin}~(l-1)\ .
\end{eqnarray}
 The explicit expressions are
as follows:
\begin{eqnarray}
 \delta\phi^i_{l(+)}&=&A^{l(+)}_{i,i_1,\cdots,i_l}
 J^{i_1}\cdots J^{i_l}\ ,
\label{phiplus}
\\
 \delta\phi^i_{l(0)}&=&\sum_{s=1}^l\epsilon_{i,i_s,j}
A^{l(0)}_{j,i_1,\cdots,i_{s-1},i_{s+1},\cdots,i_l}
 J^{i_1}\cdots J^{i_l}
=i[J^i,
A^{l(0)}_{i_1,\cdots,i_l} J^{i_1}\cdots J^{i_l}]
\ ,
\label{phizero}
\\
\delta\phi^i_{l(-)}&=&\Bigg(
\sum_{s=1}^l\delta_{i,i_s}A^{l(-)}_{i_1,\cdots,i_{s-1},i_{s+1},\cdots,i_l}
\nn\\
&&
~~~
-\frac{2}{2l-1}
\sum_{s<t}\delta_{i_s,i_t}
A^{l(-)}_{i,i_1,\cdots,i_{s-1},i_{s+1},\cdots,i_{t-1},i_{t+1},\cdots,i_l}
\Bigg)
 J^{i_1}\cdots J^{i_l}\ .
\label{phiminus}
\end{eqnarray}
Here, $A^{l(\epsilon)}_{i_1,\cdots,i_{l+\epsilon}}$
($\epsilon=0,\pm 1$) are rank-$(l+\epsilon)$
traceless symmetric tensors (spin ($l+\epsilon$)
representations) of $SO(3)$.
$\delta\phi^i_{l(0)}$ corresponds to the direction of gauge
rotation as we have seen in (\ref{gauge2}).
These modes satisfy
\begin{eqnarray}
i\epsilon^{ijk}[J^j,\delta\phi_{l(\epsilon)}^k]
=\Lambda_\epsilon\delta\phi_{l(\epsilon)}^i\ ,
\label{Lam}
\end{eqnarray}
where $\Lambda_{+}=l$, $\Lambda_0=-1$, $\Lambda_{-}=-l-1$,
and
\begin{eqnarray}
&& \{J^i,\delta\phi^i_{l(+)}\}=2
 A^{l(+)}_{i_1,\cdots,i_{l+1}}J^{i_1}\cdots
 J^{i_{l+1}}\equiv\varphi_{l(+)}\ ,
\label{varplus1}
\\
&& \{J^i,\delta\phi^i_{l(0)}\}=0\ ,
\\
&& \{J^i,\delta\phi^i_{l(-)}\}=\beta_{j,l}
A^{l(-)}_{i_1,\cdots,i_{l-1}}J^{i_1}\cdots
J^{i_{l-1}}\equiv\varphi_{l(-)}\ ,
\label{varminus1}
\end{eqnarray}
with
\begin{eqnarray}
\beta_{j,l}\equiv \frac{l^2(4j(j+1)-l^2+1)}{2(2l-1)}\ .
\label{bjl}
\end{eqnarray}
(See Appendix \ref{Ylm} for the calculation.)
Then, we obtain
\begin{eqnarray}
&&\tr\left(\{J^i,\delta\phi^i\}^2\right)\nn\\
&&=\tr\left(\varphi_{1(-)}^2\right)
\nn\\
&&
~~
+\tr\left((\varphi_{1(+)}+\varphi_{3(-)})^2\right)
+\tr\left((\varphi_{3(+)}+\varphi_{5(-)})^2\right)
+\cdots+\tr\left((\varphi_{2j-3(+)}+\varphi_{2j-1(-)})^2\right)
\nn\\
&&
~~
+\tr\left(\varphi_{2j-1(+)}^2\right)\ ,
\end{eqnarray}
where $\varphi_{l(\pm)}$ is defined in
(\ref{varplus1}) and (\ref{varminus1}).

Inserting all these into (\ref{Vphi2}), the potential in the quadratic
order with respect to the fluctuations is obtained as
\begin{eqnarray}
 V|_{\cO(\delta\phi^2)}=\sum_{k=0}^{j}V_{2k}\ ,
\end{eqnarray}
where
\begin{eqnarray}
V_{0}&=&2ga^2
\tr\left((\delta\phi^i_{1(-)})^2\right)
+\frac{\lambda a^2}{2}\tr\left(
\varphi_{1(-)}^2
\right)\ ,
\label{V0}
\\
V_{l+1}&=&\frac{ga^2}{2}\Big(
(l+1)(l-2)\tr\left((\delta\phi^i_{l(+)})^2\right)
+(l+2)(l+5)\tr\left((\delta\phi^i_{l+2(-)})^2\right)
\Big)
\label{Vl+1}
\nn\\
&&+\frac{\lambda a^2}{2}\tr\left(
(\varphi_{l(+)}+\varphi_{l+2(-)})^2
\right)\ ,~~~~(l=1,3,5,\cdots,2j-3)
\\
V_{2j}&=&ga^2
j(2j-3)\tr\left((\delta\phi^i_{2j-1(+)})^2\right)
+\frac{\lambda a^2}{2}\tr\left(
\varphi_{2j-1(+)}^2
\right)\ .
\label{V2j}
\end{eqnarray}
All terms except for $V_{2}$ are non-negative provided
$g>0$ and $\lambda>0$.
Each term can be written explicitly in terms of the
coefficients in the expansions (\ref{phiplus})--(\ref{phiminus})
by using the following formulae:
\begin{eqnarray}
 \tr\left((\delta\phi^i_{l(+)})^2\right)
&=&\alpha_{j,l}\, A^{l(+)}_{i_1,\cdots,i_{l+1}}
A^{l(+)\,i_1,\cdots,i_{l+1}}\ ,
\\
 \tr\left((\delta\phi^i_{l(-)})^2\right)
&=&\wt\alpha_{j,l}\, A^{l(-)}_{i_1,\cdots,i_{l+1}}
A^{l(-)\,i_1,\cdots,i_{l+1}}\ ,
\\
 \tr\left((\varphi_{l(+)}+\varphi_{l+2(-)})^2\right)
&=&\alpha_{j,l+1}\left(
2A^{l(+)}_{i_1,\cdots,i_{l+1}}+\beta_{j,l+2}A^{l+2(-)}_{i_1,\cdots,i_{l+1}}
\right)^2\ ,
\end{eqnarray}
where $\beta_{j,l+2}$ is as given in (\ref{bjl}) and
\begin{eqnarray}
\alpha_{j,l}=
\frac{(l!)^2}{2^l}\frac{(2j+l+1)!}{(2l+1)!(2j-l)!}
\ ,~~~
\wt\alpha_{j,l}=\frac{l^2(2l+1)}{2l-1}\alpha_{j,l}\ .
\label{wtajl}
\end{eqnarray}
(See Appendix \ref{Ylm} for the calculation.)

Then, (\ref{V0})--(\ref{V2j}) can be written as
\begin{eqnarray}
&&V_{l+1}=a^2
(\eta_{l+1},\xi_{l+1}) M_{j,l+1}
\left({\eta_{l+1}\atop\xi_{l+1}}\right)\ ,~~~
(l=1,3,\cdots,2j-3)
\nn\\
&&V_0=a^2\xi M_{\xi} \xi\ ,~~~
 V_{2j}=a^2\eta M_{\eta} \eta\ ,
\label{Vmass}
\end{eqnarray}
where
\begin{eqnarray}
&&
 \xi_{l+1}\equiv A^{l+2(-)}_{i_1,\cdots,i_{l+1}}\ ,~~
 \eta_{l+1}\equiv A^{l(+)}_{i_1,\cdots,i_{l+1}}\ ,~~
\xi\equiv \xi_{0}= A^{1(-)}\ ,~~
\eta\equiv \eta_{2j}=A^{2j-1(+)}_{i_1,\cdots,i_{2j}}\ ,
\end{eqnarray}
and
\begin{eqnarray}
M_{j,l+1}&=&\mat{
\frac{g}{2}(l+1)(l-2)\alpha_{j,l}+2\lambda\alpha_{j,l+1},
\lambda\alpha_{j,l+1}\beta_{j,l+2},
\lambda\alpha_{j,l+1}\beta_{j,l+2},
\frac{g}{2}(l+2)(l+5)\wt\alpha_{j,l+2}
+\frac{\lambda}{2}\alpha_{j,l+1}\beta_{j,l+2}^2
}\ ,
\label{Mjl+1}
\nn\\
\\
M_\xi
&=&
2g\wt\alpha_{j,1}
+\frac{\lambda}{2}\alpha_{j,0}\beta_{j,1}^2
=2j(j+1)(2j+1)
\left(g+\lambda j(j+1)
\right)\ ,
\label{Mxi}
\\
M_\eta
&=&
g j (2j-3)\alpha_{j,2j-1}+2\lambda\alpha_{j,2j}
=\frac{((2j)!)^2}{2^{2j-1}}\left(
g(2j-3)+
\lambda\right) \ .
\label{Meta}
\end{eqnarray}
Here, the indices $i_1,\cdots,i_{l+1}$ for $\xi_{l+1}$
and $\eta_{l+1}$ are suppressed, and contracted appropriately
in the expression (\ref{Vmass}).

In particular, for $j=1$, (\ref{Mxi}) and (\ref{Meta}) are
\begin{eqnarray}
 M_\xi
=12(2\lambda+g)\ ,~~~
 M_\eta=2(\lambda-g)\ ,
\end{eqnarray}
which reproduces (\ref{j=1pot}).

As we have observed above, the instability can possibly
occur only in $V_2$ for $j>1$.
The matrix (\ref{Mjl+1}) with $l=1$ is
\begin{eqnarray}
M_{j,2}
&=&
j(j+1)(2j+1)\wt M_{j,2}\ ,
\end{eqnarray}
where
\begin{eqnarray}
\wt M_{j,2}=
\mat{\frac{-5g+(2j-1)(2j+3)\lambda}{15},
\frac{3(j-1)(j+2)(2j-1)(2j+3)\lambda}{25},
\frac{3(j-1)(j+2)(2j-1)(2j+3)\lambda}{25},
\frac{27(j-1)(j+2)(2j-1)(2j+3)(15g+2(j-1)(j+2)\lambda)}{250}
}\ .
\end{eqnarray}
The trace of this matrix is always positive and
the determinant is
\begin{eqnarray}
 \det\wt M_{j,2}
=\frac{9}{50}g(j-1)(j+2)(2j-1)(2j+3)
(-3g+(2j^2+2j-1)\lambda)\ .
\end{eqnarray}
Therefore, if
\begin{eqnarray}
 \lambda>\frac{3}{2j^2+2j-1}g\ ,
\label{ljg}
\end{eqnarray}
we have $\det\wt M_{j,2}>0$ and
there is no negative eigenvalue that causes the instability.
The condition (\ref{ljg}) is always satisfied when
(\ref{lng}) is satisfied.

\section{Summary and discussion}
\label{sec5}

We have argued that the electric theory in \S \ref{eleth} and
the magnetic theory in \S \ref{magth} are dual to each other, because
they are obtained as low-energy effective theories of the O3-\AD3
systems related by the S-duality in type IIB string theory.
The electric theory is conjectured to be a confining theory
and the global $SO(6)$ symmetry is conjectured to be dynamically broken
to the $SO(4)$ subgroup via the fermion bilinear condensate (\ref{TrQQ}).
These properties can be understood in terms of the magnetic theory as a
consequence of the condensation of tachyonic scalar fields, which is
consistent with the scenario of the dual Meissner mechanism of the
confinement.

There are some subtleties which should be addressed to make the argument
more accurate. One important problem is to see what happens
when we try to take the decoupling limit in the magnetic theory.
Since our magnetic theory is an asymptotically non-free theory,
we have treated it as a low-energy effective theory of the \Omt3-\AD3
systems in string theory without trying to take the $l_s\ra 0$ limit.
For this reason, we can only trust the analysis at low energies,
and we are not able to cover the whole energy region.
In any case, since both the electric and magnetic theories are strongly
coupled around the dynamical scale of the systems, the duality
is not powerful enough to make a quantitatively accurate analysis.
Because of this limitation, we had to rely on a toy model to analyze
the system. However, it is evidently desirable to derive everything
without making any speculative assumptions.

It would be interesting to consider the generalization and application
of this idea.
The simplest generalization is to use the \Om3-\AD3 system to construct
a non-supersymmetric $SO(2n)$ gauge theory. The field content is
similar to our magnetic theory (II), although the gauge group
is different. Since the \Om3-plane is
self-dual, this $SO(2n)$ gauge theory is predicted to be self-dual.
We have not explored the consequence of the duality in this system.
Another interesting direction would be to consider the holographic dual
of the system. As mentioned in \S \ref{first},
our system is related to the $\cN=4$ SYM at the leading order
in the $1/n$ expansion. Therefore, the holographic dual of this system
is type IIB string theory in the $AdS_5\times S^5$ background with the $\Z_2$
orientifold action identifying the antipodal points of $S^5$ at the
leading order in the $1/n$ expansion.
In order to see the effect of SUSY breaking, $1/n$ corrections should
be taken into account. The systems considered in this paper may be useful
for the investigation of the $1/n$ corrections in the context
of gauge/string duality.

\section*{Acknowledgements}

I would like to thank Adi Armoni, Oren Bergman, Kentaro Hori,
Takeshi Morita, Yutaka Ookouchi, Soo-Jong Rey, Yuji Tachikawa, Piljin Yi
and my colleagues at the Kavli Institute for the Physics and Mathematics
of the Universe (Kavli IPMU) for helpful discussions.
I am especially grateful to Yutaka Ookouchi for encouragement and
pleasant collaboration at the early stage of this work.
This work was supported in part by Grants-in-Aid for Young Scientists
(B) (No. 21740173), Scientific Research (C) (No. 24540259) and Creative
Scientific Research (No. 19GS0219) from the Japan Society for the Promotion
of Science (JSPS), and also by World Premier International
Research Center Initiative (WPI Initiative), MEXT, Japan. 

\appendix

\section{Notation for $USp(2n)$}
\label{USp}

$USp(2n)$ is defined as the set of the elements $g\in SU(2n)$ satisfying
\begin{eqnarray}
g^T Jg=J\ ,
\end{eqnarray}
where $J$ is the antisymmetric invariant tensor
defined as 
\begin{eqnarray}
J\equiv\mat{0,1_n,-1_n,0}
\end{eqnarray}
with the $n\times n$ unit matrix $1_n$.

The Lie algebra associated with the $USp(2n)$ group is the set of
$2n\times 2n$ anti-Hermitian matrices $X$ satisfying
\begin{eqnarray}
X^T J+JX =0\ ,
\label{XJJX}
\end{eqnarray}
which is equivalent to the condition that $JX$ is a symmetric matrix.
We can use $J=(J_{ab})=(J^{ab})$ to raise or lower the indices for the matrix
$X=(X^a_{~b})$ such as
\begin{eqnarray}
X_{ab}=J_{ac}X^c_{~b}\ .
\end{eqnarray}
Then, (\ref{XJJX}) can be written as $X_{ab}=X_{ba}$.
The elements of the representation space of the
rank-2 symmetric and antisymmetric tensor representations
($\sym$ and $\asym$) have the index structures
$X_{ab}=X_{ba}$ and $X_{ab}=-X_{ba}$, respectively.
The antisymmetric tensor representation $\asym$
is a reducible representation that can be decomposed to a singlet
component proportional to $J_{ab}$ and
its orthogonal complement satisfying
\begin{eqnarray}
 X_{ab}J^{ab}=0\ .
\end{eqnarray}
In our matrix notation for the fields $A_\mu$, $Q^i$ and $\Phi^I$
listed in Table \ref{ele}, the gauge indices are assumed to be
like $(A_\mu)^{a}_{~b}$, $(Q^i)^{a}_{~b}$ and $(\Phi^I)^{a}_{~b}$.
They are subject to condition (\ref{JAJQJP}).

\section{The Explicit Form of $\Sigma^I_{ij}$}
\label{Sigma}

$\Sigma^I$ used in (\ref{Lele}), (\ref{Lmag1})  and (\ref{Lmag2}) 
is represented as
\begin{eqnarray}
&&
 \Sigma^1 =\mat{0,\sigma_1,-\sigma_1,0}\ ,~~
 \Sigma^2 =\mat{0,-\sigma_3,\sigma_3,0}\ ,~~
 \Sigma^3 =\mat{i\sigma_2,0,0,i\sigma_2}\ ,
\nn\\
&&
 \Sigma^4 =\mat{0,\sigma_2,\sigma_2,0}\ ,~~
 \Sigma^5 =\mat{0,-i1_2,i1_2,0}\ ,~~
 \Sigma^6 =\mat{-\sigma_2,0,0,\sigma_2}\ .
\end{eqnarray}
They satisfy
\begin{eqnarray}
(\Sigma^{I\dag} \Sigma^{J}+\Sigma^{J\dag} \Sigma^{I})^i_{~j}=
2\delta^{IJ}\delta^i_j\
\end{eqnarray}
and
\begin{eqnarray}
 \Gamma^I=\mat{0,\Sigma^I,\Sigma^{I\dag},0}
\end{eqnarray}
are the $SO(6)$ gamma matrices.

The generators of $SO(6)$ in the spinor representation are given by
\begin{eqnarray}
\ol\Sigma^{IJ}\equiv
 \frac{1}{4i}(\Sigma^{I\dag}\Sigma^J-\Sigma^{J\dag}\Sigma^I)\ .
\end{eqnarray}
The explicit forms for the generators of the $SO(3)\times SO(3)$
subgroup in the spinor representation are given by 
\begin{eqnarray}
&&\ol\Sigma^{12}=\frac{1}{2}
\mat{\sigma_2,0,0,\sigma_2}\ ,~~
\ol\Sigma^{23}=\frac{1}{2}
\mat{0,-i\sigma_1,i\sigma_1,0}\ ,~~
\ol\Sigma^{31}=\frac{1}{2}
\mat{0,i\sigma_3,-i\sigma_3,0}\ ,
\nn\\
&&\ol\Sigma^{45}=\frac{1}{2}
\mat{\sigma_2,0,0,-\sigma_2}\ ,~~
\ol\Sigma^{56}=\frac{1}{2}
\mat{0,-\sigma_2,-\sigma_2,0}\ ,~~
\ol\Sigma^{64}=\frac{1}{2}
\mat{0,i1_2,-i1_2,0}\ .
\nn\\
\label{SO3SO3}
\end{eqnarray}
Note that the matrices in (\ref{SO3SO3}) are pure imaginary
antisymmetric matrices and hence they form a basis of the generators of
the $SO(4)$ subgroup of $SU(4)$.

\section{Useful Formulae for $V|_{\cO(\delta\phi^2)}$}
\label{Junk}

Here, we summarize the formulae that are used
to obtain (\ref{Vphi2}).
We only consider quadratic terms with respect to the fluctuation
$\delta\phi^I$ in (\ref{fluc}).
\begin{eqnarray}
\tr(\phi^I\phi^I)
&=&
a^2j(j+1)(2j+1)
+2 a \tr(J^i\delta\phi^i)
+\tr(\delta\phi^I\delta\phi^I)\ ,
\label{phiIphiI}
\end{eqnarray}
\begin{eqnarray}
\tr([\phi^I,\phi^J]^2)
&=&
-a^42j(j+1)(2j+1)
-8a^3\tr(J^j\delta\phi^j)
\nn\\
&&
+2a^2i\epsilon^{ijk}\tr([J^k,\delta\phi^i]\delta\phi^j)
+2a^2\tr([J^i,\delta\phi^m][J^i,\delta\phi^m])
\nn\\
&&
+2a^2\epsilon^{ijk}\epsilon^{i'j'k}
\tr([J^i,\delta\phi^j][J^{i'},\delta\phi^{j'}])
+\cO(\delta\phi^3)\ ,
\label{phiIphiJ2}
\end{eqnarray}
\begin{eqnarray}
\tr((\phi^I\phi^I)^2)
&=&
a^4j^2(j+1)^2(2j+1)
+4a^3j(j+1)\tr(J^j\delta\phi^j)
\nn\\
&&
+2a^2j(j+1)\tr(\delta\phi^J\delta\phi^J)
+a^2\tr(\{J^i,\delta\phi^i\}\{J^j,\delta\phi^j\})
+\cO(\delta\phi^3) \ .
\label{phiIphiI2}
\end{eqnarray}
Here, the ranges of the indices are $I,J=1\sim 6$; $i,j,k=1\sim 3$ and
$m=4\sim 6$.

Inserting (\ref{phiIphiI}), (\ref{phiIphiJ2}) and  (\ref{phiIphiI2})
into the potential (\ref{Vphi}), we see that the $\cO(\delta\phi^1)$
terms vanish and the $\cO(\delta\phi^0)$ terms recover the result
in (\ref{oneS2}) if we choose $a$ as in (\ref{a}).
The $\cO(\delta\phi^2)$ terms are 
\begin{eqnarray}
 V|_{\cO(\delta\phi^2)}
&=&
-\frac{\mu^2}{2}
\tr(\delta\phi^I\delta\phi^I)
\nn\\
&&
-\frac{ga^2}{2}
\tr\left([J^i,\delta\phi^m]^2
+
i\epsilon^{ijk}[J^j,\delta\phi^k]
\left(\delta\phi^i-i\epsilon^{ij'k'}
[J^{j'},\delta\phi^{k'}]\right)
\right)
\nn\\
&&
+\frac{\lambda a^2}{2}\tr
\left(2j(j+1)\delta\phi^I\delta\phi^I
+\{J^i,\delta\phi^i\}^2
\right)\ .
\label{c4}
\end{eqnarray}
Using the relation (\ref{a}),
(\ref{c4}) can be written as
\begin{eqnarray}
 V|_{\cO(\delta\phi^2)}
&=&
-ga^2
\tr\left(\delta\phi^m\delta\phi^m
+\half[J^i,\delta\phi^m]^2\right)
\nn\\
&&
-\frac{ga^2}{2}
\tr\bigg(
\left(\delta\phi^i+i\epsilon^{ijk}
[J^{j},\delta\phi^{k}]\right)
\left(2\delta\phi^i-i\epsilon^{ij'k'}
[J^{j'},\delta\phi^{k'}]\right)
\bigg)
\nn\\
&&
+\frac{\lambda a^2}{2}\tr
\left(\{J^i,\delta\phi^i\}^2
\right)\ .
\end{eqnarray}

\section{Useful Formulae for $J^i$}
\label{Ylm}

Let $J^i$ ($i=1,2,3$) be the generators of
the $su(2)$ algebra in the spin-$j$ representation, which satisfy
\begin{eqnarray}
 [J^i,J^j]=i\epsilon^{ijk}J^k\ .
\label{su2-2}
\end{eqnarray}
These $J^i$ are $(2j+1)\times (2j+1)$ Hermitian matrices.
For integer spin $j\in \Z$, they can be represented as pure imaginary
antisymmetric matrices.

The $su(2)$ algebra (\ref{su2-2}) can also be written as
\begin{eqnarray}
 [J^+,J^-]=2J^3\ ,~~[J^3,J^\pm]=\pm J^\pm\ ,
\end{eqnarray}
where $J^\pm=J^1+iJ^2$.
The spin-$j$ representation is characterized by the lowest-weight state
$\ket{0}$ satisfying
\begin{eqnarray}
J^- \ket{0}=0\ ,~~J^3 \ket{0}=-j\ket{0}\ .
\end{eqnarray}
The orthonormal basis of the representation space is given by
\begin{eqnarray}
 \ket{k}\equiv \sqrt{\frac{(2j-k)!}{(2j)!\,k!}}(J^+)^k\ket{0}\ ,
~~~(k=0,1,\cdots,2j)
\end{eqnarray}
which satisfy the relation
\begin{eqnarray}
 (J^+)^l\ket{k}=\sqrt{\frac{(2j-k)!\,(k+l)!}{(2j-k-l)!\, k!}}\ket{k+l}\ .
\end{eqnarray}
This implies
\begin{eqnarray}
\tr\left((J^-)^l(J^+)^l\right)
&=&\sum_{k=0}^{2j-l}||(J^+)^l\ket{k}||^2
\nn\\
&=&\sum_{k=0}^{2j-l}
\frac{(2j-k)!\,(k+l)!}{(2j-k-l)!\, k!}
\nn\\
&=&(l!)^2{}_{2j+l+1}C_{2l+1}
=\frac{(l!)^2(2j+l+1)!}{(2l+1)!\,(2j-l)!}\ .
\label{trJlJl}
\end{eqnarray}
It is not difficult to show
\begin{eqnarray}
 \tr(J^iJ^j)=\frac{1}{3}j(j+1)(2j+1)\delta^{ij}\ ,
\label{JJ}
\end{eqnarray}
\begin{eqnarray}
 \tr(J^iJ^jJ^kJ^l)
=a\delta^{ij}\delta^{kl}+a\delta^{il}\delta^{jk}
+b\delta^{ik}\delta^{jl}\ ,
\label{JJJJ}
\end{eqnarray}
where
\begin{eqnarray}
a&=&\frac{1}{15}j(j+1)(2j+1)\left(j(j+1)+\half\right)\ ,
\\
b&=&\frac{1}{15}j(j+1)(2j+1)\left(j(j+1)-2\right)\ .
\end{eqnarray}

Using (\ref{trJlJl}), we obtain
\begin{eqnarray}
\tr\left(
(A_{i_1,\cdots,i_l}J^{i_1}\cdots J^{i_l})
(B_{j_1,\cdots,j_l}J^{j_1}\cdots J^{j_l})
\right)=\alpha_{j,l}A^{i_1,\cdots,i_l}B_{i_1,\cdots,i_l}\ ,
\end{eqnarray}
where $A_{i_1,\cdots,i_l}$ and
$B_{i_1,\cdots,i_l}$ are rank-$l$
traceless symmetric tensors of $SO(3)$, and
\begin{eqnarray}
 \alpha_{j,l}=\frac{(l!)^2}{2^l}\,{}_{2j+l+1}C_{2l+1}
=\frac{(l!)^2}{2^l}\frac{(2j+l+1)!}{(2l+1)!(2j-l)!}\ .
\end{eqnarray}

The following formulae are also useful:
\begin{eqnarray}
&& J^a J^a=j(j+1)1_{2j+1}\ ,
\label{JaJa}
\\
&& J^a J^i J^a=(j(j+1)-1)J^i\ ,
\\
&&
J^a J^iJ^j J^a=(j(j+1)-3)J^iJ^j+i\epsilon^{ijk} J^k
+\delta^{ij}j(j+1) 1_{2j+1}\ .
\end{eqnarray}

For
\begin{eqnarray}
A\equiv A_{i_1,\cdots,i_l}J^{i_1}\cdots J^{i_l}\ ,
\end{eqnarray}
with a traceless symmetric tensor $A_{i_1,\cdots,i_l}$,
we can show
\begin{eqnarray}
 J^a A J^a = \left(j(j+1)-\half l(l+1)\right) A
\end{eqnarray}
by induction. This formula can also be checked by using the relation
(\ref{JaJa}) and another useful formula
\begin{eqnarray}
 [J^a,[J^a,A]]=l(l+1)A\ .
\end{eqnarray}

Using these relations, we obtain
\begin{eqnarray}
&& \{J^i,\delta\phi^i_{l(+)}\}=2
 A^{l(+)}_{i_1,\cdots,i_{l+1}}J^{i_1}\cdots
 J^{i_{l+1}}\equiv\varphi_{l(+)}\ ,
\\
&& \{J^i,\delta\phi^i_{l(0)}\}=0\ ,
\\
&& \{J^i,\delta\phi^i_{l(-)}\}=\beta_{j,l}
A^{l(-)}_{i_1,\cdots,i_{l-1}}J^{i_1}\cdots
J^{i_{l-1}}\equiv\varphi_{l(-)}\ ,
\end{eqnarray}
where
\begin{eqnarray}
\beta_{j,l}= \frac{l^2(4j(j+1)-l^2+1)}{2(2l-1)}\ ,
\end{eqnarray}
for $\delta\phi^i_{l(\epsilon)}$ defined in
(\ref{phiplus})--(\ref{phiminus}).

%

\end{document}